\documentclass[preprint,3p,authoryear,a4paper]{elsarticle}
\usepackage{graphicx}
\usepackage{epstopdf}
\usepackage{color}
\usepackage{dsfont}
\usepackage{amsmath}
\usepackage{amssymb}
\usepackage{amsfonts}
\usepackage{amsbsy}
\usepackage{caption}
\usepackage{subcaption}
\usepackage{amsthm}
\usepackage{array}
\usepackage{booktabs} 
\usepackage{verbatim}
\usepackage[english]{babel}
\usepackage{natbib}
\usepackage{ulem}
\usepackage{url}
\usepackage{multirow}
\usepackage{subcaption}
\usepackage{float}
\usepackage{numprint}

\renewcommand\appendix{\par
\setcounter{section}{0}
\setcounter{subsection}{0}
\setcounter{table}{0}
\setcounter{figure}{0}
\gdef\thetable{\Alph{table}}
\gdef\thefigure{\Alph{figure}}
\gdef\thesection{\Alph{section}}
\setcounter{section}{0}}

\newtheorem{theorem}{Theorem}[section]

\newtheorem{lemma}[theorem]{Lemma}

\newtheorem{remark}{Remark}[section]
\numberwithin{equation}{section}

\newenvironment{arcitem}{
\begin{list}{---}{
\topsep=1pt
\itemsep=1pt 
\parsep=0pt 
\leftmargin=19pt 
}}
{\end{list}}

\newcounter{arclist}

\newcounter{arcenum}

\begin{document}

\normalem

\begin{frontmatter}

\title{Stochastic loss reserving with mixture density neural networks}

% Stochastic loss reserving with neural networks: \\ distribution selection with the MDN and ResMDN
% MDN and ResMDN: Stochastic loss reserving via neural networks with distribution selection
% Stochastic loss reserving with neural networks: a Mixture Density network approach.
% Stochastic loss reserving with mixture density neural networks

\author[UNSW]{Muhammed Taher Al-Mudafer\corref{cor}}
\ead{muhammed.almudafer97@gmail.com}

\author[UMelb]{Benjamin Avanzi}
\ead{b.avanzi@unimelb.edu.au}

\author[UNSW]{Greg Taylor}
\ead{gregory.taylor@unsw.edu.au}

\author[UNSW]{Bernard Wong}
\ead{bernard.wong@unsw.edu.au}

\cortext[cor]{Corresponding author. }

\address[UNSW]{School of Risk and Actuarial Studies, UNSW Australia Business School, UNSW Sydney NSW 2052, Australia}
\address[UMelb]{Centre for Actuarial Studies, Department of Economics, University of Melbourne VIC 3010, Australia}

\begin{abstract}

In recent years, new techniques based on artificial intelligence and machine learning in particular have been making a revolution in the work of actuaries, including in loss reserving. A particularly promising technique is that of neural networks, which have been shown to offer a versatile, flexible and accurate approach to loss reserving. However, applications of neural networks in loss reserving to date have been primarily focused on the (important) problem of fitting accurate central estimates of the outstanding claims. In practice, properties regarding the variability of outstanding claims are equally important (e.g., quantiles for regulatory purposes). 

In this paper we fill this gap by applying a Mixture Density Network (``MDN'') to loss reserving. The approach combines a neural network architecture with a mixture Gaussian distribution to achieve simultaneously an accurate central estimate along with flexible distributional choice. Model fitting is done using a rolling-origin approach. Our approach consistently outperforms the classical over-dispersed model both for central estimates and quantiles of interest, when applied to a wide range of simulated environments of various complexity and specifications.

We further extend the MDN approach by proposing two extensions. Firstly, we present a hybrid GLM-MDN approach called ``ResMDN``. This hybrid approach balances the tractability and ease of understanding of a traditional GLM model on one hand, with the additional accuracy and distributional flexibility provided by the MDN on the other. We show that it can successfully improve the errors of the baseline ccODP, although there is generally a loss of performance when compared to the MDN in the examples we considered. Secondly, we allow for explicit projection constraints, so that actuarial judgement can be directly incorporated in the modelling process. 

Throughout, we focus on aggregate loss triangles, and show that our methodologies are tractable, and that they out-perform traditional approaches even with relatively limited amounts of data. We use both simulated data---to validate properties, and real data---to illustrate and ascertain practicality of the approaches. 
\end{abstract}

\begin{keyword} Loss reserving, neural network, Mixture Density Network, Distributional Forecasting, machine learning

%C51 \sep % C5	Econometric Modeling C51	Model Construction and Estimation
C53 \sep % C5	Econometric Modeling C53	Forecasting and Prediction Methods • Simulation Methods 
%C55 \sep % C5	Econometric Modeling C55	Large Data Sets: Modeling and Analysis
G22 \sep % G2	Financial Institutions and Services G22	Insurance • Insurance Companies • Actuarial Studies

MSC classes: 
%60G51 \sep % Processes with independent increments
%93E20 \sep % Optimal stochastic control
91G70 \sep 	%Statistical methods; risk measures [See also 62P05, 62P20] in Actuarial science and mathematical finance
91G60 \sep 	%Numerical methods (including Monte Carlo methods) in Actuarial science and mathematical finance
62P05 %\sep 	%Applications of statistics to actuarial sciences and financial mathematics
%62H12 %\sep 	%Estimation in multivariate analysis
%91B30 %\sep % Risk theory, insurance

\end{keyword}
\end{frontmatter}
{\centering \large}

\section{Introduction}\label{sec:intro}

\subsection{Background}

Forecasting outstanding claims (``OSC'') is essential in insurance for effective loss reserving, liability reporting and pricing. In addition to producing accurate central estimates, forecasting the \emph{distribution} of future claims accurately is important in allocating capital which satisfies regulatory requirements. In Australia, the calculation of regulated risk margins involves determining quantiles such as the 75\% percentile. In Europe, Solvency II regulations require calculations as far as a 99.5\% percentile. Having an enhanced understanding of the distribution of OSC is beneficial beyond regulatory obligations, of course. With \emph{point forecasts} of OSC being the predominant focus on the literature, this paper focuses on \emph{distributional} forecasting of outstanding claims using neural networks.

Machine learning models, especially neural networks (``NN''), have gained momentum in the actuarial field in the past few years. Neural networks have stood out for providing ‘state of the art results’ \citep*{RoRi2019}. Their earliest use in reserving, to our knowledge, goes to \citet*{Mu2006}. Since 2018, this field has accelerated, with recent NN applications to reserving showcasing its accuracy and versatility. \citet*{Ku2019, GaRiWu2020} use NNs to learn claim development trends from multiple lines of business simultaneously. \citet*{WuMe2019} develop a GLM-NN hybrid model, which is used successfully by \citet*{GaRiWu2020,Po2019,Ga2020}. The NNs ability to handle large granular datasets has allowed individual claims modelling to flourish, with recent success demonstrated by \citet*{Ku2020,DeLiWu2020,Ga2020}.

\subsection{Motivation and contributions}

Despite the potential shown by NNs, several gaps in their current implementation can be observed, in particular with aggregate data. Firstly, most of the neural network loss reserving literature focuses on obtaining accurate central estimates. However, providing accurate \emph{distributional} forecasts for outstanding claims is essential for optimising capital allocation, reporting liabilities with more accurate risk margins, and allowing profit margins that suit the company's risk appetite to be set more accurately. Secondly, there is a lack of an \emph{explicit model selection framework}. The performance of neural networks is heavily dependent on its design, therefore having no explicit selection procedure risks choosing a model with significantly reduced accuracy. Furthermore, model building becomes more reliant on expert input, hindering their \emph{applicability in practice}. This paper aims to address those issues, as developed below.

\subsubsection{Distributional forecasting with Neural Networks using the MDN}
Probabilistic forecasts of outstanding claims are essential for capital allocation and optimising the risk margins set when reporting liabilities and pricing products. In this paper, we explain how such forecasts can be obtained with the Mixture Density Network (``MDN'') in a flexible way, working with aggregate loss triangles. The MDN is a design developed by \citet*{Bi1994} \citep*[see also][]{Bi06}, with applications found in many fields, such as financial modelling \citep*{OrNe1996}, acoustics \citep*{ZeSe2014} and electrical engineering \citep*{VoFeMo2018}. The essential idea is that the parameters of the distribution will be estimated by the NN architecture, allowing for heterogeneity. Related works in the (non mixture) Gaussian setting can be found in \citet{NiWe94} and \citet{LaPrBl17}. The MDN assumes that the target output follows a mixture (typically Gaussian) distribution, allowing effectively a flexible distribution to be fit to the data. In this paper, we will assume that incremental claims follow a mixture Gaussian or mixture Log-Gaussian distribution.

A special case of an MDN was used by \citet{Ku2020} who mixed a shifted lognormal with a degenerate distribution to represent (positive) and zero cashflows for individual claims, respectively. In contrast, in our application we utilise a mixture of Gaussian distributions, whereby the number of components as well as all parameters are kept flexible.

In loss reserving, previous literature has fitted members of the exponential family using NNs. For instance, \citet*{GaRiWu2020} considered aggregate claims, and \citet{DeLiWu2020,Ga2020} considered individual claims. However, they typically focused on the central estimate rather than distributional properties. 

A typical approach is to utilise the NN to replace the linear component in a GLM setup   \citep*[see also][for additional discussions]{DeTr2019}, including the GLM-based assumption of a homogeneous mean-variance relationship. Such an assumption is not as flexible for the modelling of volatility, as compared to the mixture Gaussian fit by the MDN, which allows for heterogeneity for \emph{each} random variable. In addition, the mixture Gaussian, given sufficient components, can approximate any distribution  within a desired accuracy \citep[see][for details]{NgMc2019}. As the MDN considers a wide range of distributions of varying volatility and shape, the training and fitting process of the network is akin to distribution selection. Another benefit of MDNs is that the central estimates can be derived directly from the fitted mixture Gaussian parameters, which means that  location and shape are fitted simultaneously under the single model. 

In addition to analysing central estimates, we use qualitative and quantitative measures to assess the distributional and quantile accuracy of the fitted MDN forecasts; see Section \ref{sec:metrics}. Overall, we show that the MDN's flexibility yields more accurate probabilistic forecasts when compared to the cross-classified over-dispersed Poisson (``ccODP''), in both simulated and real data; see also Section \ref{S_motdata}. 

\subsubsection{Model calibration and selection}
Neural networks are highly flexible in their design. Since different designs produce varying results, it is vital to have a clear methodology for testing between these different model designs. 

Commonly, the data is partitioned into training, validation and testing sets. %can mention need for sequential partition
The network is trained on the training set until the validation loss is minimised, then projected onto the testing set. The model producing the lowest test error is preferred.

To test between different models, a loss triangle must be split into training, validation and testing sets. \citet*{Ta2000} and \citet*{BeBe2012} explore two popular methods; fixed origin and rolling origin, used to partition sequential data. \citet*{BaRi2020} recently apply the rolling origin method to loss triangles, in order to compare between different traditional reserving models.

This paper contributes by performing the rolling origin data partition exclusively within the aggregate loss triangle, using it to select neural network designs. Taking the latest calendar periods for testing has been done by \citet*{Ku2019,RaAlNu2019}, which was facilitated by simultaneously training the NN on multiple triangles. This paper performs model testing, selection and fitting using only one triangle at a time. Furthermore, the neural network model testing framework implemented in this paper outlines a validation set, which is essential in training NNs.

Furthermore, this paper contributes by implementing a model searching and selection algorithm to loss triangle reserving, which methodically searches and selects different network features such as the number of layers, nodes, and components. This algorithm extends the methodology implemented by \citet*{GaRiWu2020} by also searching for the best-performing regularisation coefficients. This algorithm also makes NN design selection more methodical, requiring less expert input and assisting the popularisation of NNs in practice.

\subsubsection{Acceptability of results in practice: ResMDN, and projection constraints}

While the current neural network applications to loss reserving have shown their accuracy and flexibility in modelling, there are practical and technical obstacles which have hindered the acceptance and use of neural networks by the actuarial community \citep*{RoRi2019, WuMe2019}. %Analysing and fixing these shortcomings and expanding the implementation of neural networks in the literature can increase their reliability and practicality, which will in turn encourage the NNs modelling power to be applied by in practical loss reserving, adding more models available to the actuary and improving the reserving process.

Producing interpretable forecasts is important for justifying business decisions to stakeholders. The neural network’s lack of interpretability has contributed to its lack of acceptance by actuaries. \citet*{WuMe2019} tackle this by adopting the resnet design from \citet{HeZhReSu16} to form the GLM-NN hybrid CANN (Combined Actuarial Neural Network) architecture. 

In this paper, we adapt the MDN to the above approach, to create the ResMDN. While \citet*{GaRiWu2020}, \citet{Po2019} and \citet{RiWu2021} have adopted a similar methodology in loss reserving, the ResMDN provides the additional distributional flexibility of a (heterogeneous) mixture Gaussian distribution assumption. It boosts all parameters of the mixture Gaussian simultaneously and directly within one network, while maintaining a fixed GLM initialisation for interpretability. 

We show that the ResMDN can successfully boost the embedded GLM, the ccODP, improving structural deficiencies in both mean and volatility estimates where visible, while maintaining the interpretable GLM backbone in its forecasts.

As a further consideration, the NN's black box modelling can cause long range forecasts to become unstable \citep{RoRi2019}. In other words, it may not be adequate to project the behaviour of the highly flexible function fit to the upper triangle to the lower triangle. In this paper, central estimate projections are able to be explicitly constrained by the actuary, forcing the neural network to only fit functions which produce reasonable mean forecasts. This framework thus also allows actuarial judgement to be explicitly incorporated into the modelling process.

\subsubsection{Aggregate data in loss triangles} \label{S_motdata}

Recent applications of machine learning to loss reserving have been mainly focused on more granular claims datasets, such as individual claims data. This article, however, works exclusively with aggregate loss triangles. Developing a neural network model that is applied exclusively to loss triangles is highly relevant, as it enhances comparability and interpretability of results (as opposed to traditional reserving), and some insurers may still have limited data, preventing them from applying individual loss reserving methodologies. Furthermore, it is not obvious that machine learning methods can replicate their documented accuracy when presented with limited data, which is an interesting research question in itself.

In this paper, we demonstrate how neural networks can improve on traditional models, even with limited triangular data:
\begin{arcitem}
    \item \emph{Data scarcity:} The MDN was applied by Kuo (2020), but it was used in Individual Claims modelling. The current paper shows that the MDN also finds success when applied to loss triangles.
    \item \emph{Model selection:} Rossouw \& Richman (2019) outline a fixed origin data partition (see Bergmeir \& Benitez (2012) for more detail), but use more granular data, hence this methodology hasn't been tested on a sparse loss triangle. While Gabrielli et al (2020) works with loss triangles, individual claims data are used to partition into training and validation sets. Balona \& Richman (2020) recently apply the rolling origin model validation methodology (see Bergmeir \& Benitez (2012) for more details) exclusively to loss triangles. However, as only non-machine learning models were tested, no validation set was constructed in the paper.
    
    With the rolling origin methodology and model searching algorithm, the neural network designs chosen in this paper produce robust, smooth, and accurate central and distributional forecasts, showcasing their combined effectiveness. Given that neural networks can perform unreliably with small datasets, these results show that the rolling origin partition is feasible for loss triangles of a certain size (40 $\times$ 40 triangles were used in this paper), which should encourage further neural network modelling with loss triangles.
    
    \item \emph{Data variety:} Additionally, for NNs to prove their practicality and reliability, they must provide accurate results in a variety of different triangles (of instance, in a range of claim development situations which would invalidate the chain ladder assumptions). The current paper contributes to the literature by testing the MDN on a variety of environments of varying complexity and specifications, using both simulated and real data. Some of the environments tested are derived in concept from triangles used by \cite{HaGaJa2017, GaRiWu2020}. The MDN achieved superior probabilistic forecasts relative to the ccODP in all environments tested (see Section \ref{sec:Data} for details).
\end{arcitem}

\subsection{Structure of the paper}
Section \ref{sec:ModelDesc} provides a detailed overview of the models used in this paper; the MDN, ResMDN and the benchmark ccODP models. Section \ref{sec:ModelDev} outlines the model development methodologies, including rolling origin validation and the hyper-parameter selection algorithm. The section concludes with an overview of the environments used. Section \ref{sec:training} provides details into the network training procedure and evaluation metrics. The standard MDN's results are analysed in Section \ref{sec:Results}, with practical considerations, including the ResMDN, explored in Section \ref{sec:practical}. Section \ref{sec:conclusion} concludes.

\section{Description of the ccODP, MDN and ResMDN models}
\label{sec:ModelDesc}
\subsection{Notation}

Let us first introduce some basic notation. Let
\begin{itemize}
\item $\Phi(x | \mu ,\sigma) = P(Z \leq x)$ be the distribution function of a normal distribution with mean $\mu$ and standard deviation $\sigma$, that is, $Z \sim N(\mu, \sigma)$;
\item $ d\Phi(x| \mu, \sigma)=\phi(x | \mu, \sigma) dx$, that is, $\phi(x | \mu, \sigma) $ is the probability density function of $Z$;
\item $X_{i,j}$ be the incremental claims paid in accident period $i$ and Development period $j$;
\item  $\hat{X}_{i,j} $ be a random variable which a model predicts to match the distribution of $X_{i,j}$; 
\item AQ and DQ be abbreviations for accident quarter and development quarter, respectively.
\end{itemize}

\subsection{GLM: Cross-Classified Over-Dispersed Poisson (ccODP) }
\label{sec:ccODP}
The benchmark model used in this paper is the Cross-Classified Over-Dispersed Poisson model (``ccODP''), in line with the existing NN loss reserving literature \citep*{Ku2019,GaRiWu2020,Ga2019,DeLiWu2020,Ku2020,Wu2018}. The ccODP model assumes that incremental claims, $X_{i,j}$, follow a Cross-Classified Over-Dispersed Poisson distribution
\begin{equation}
\frac{\hat{X}_{i,j}}{D} \sim \text{Poi} \bigg(\frac{A_iB_j}{D} \bigg), 
\text{ where }E[\hat{X}_{i,j}] = A_iB_j\text{ and }Var(\hat{X}_{i,j}) = D A_iB_j.
\label{eqn:ccODP}
\end{equation}
The Cross-Classified structure of the model, as well as the assumption of a constant $D$, leads it to produce mean estimates that are identical to the Chain Ladder. Hence, the ccODP's strengths and weaknesses in central estimate accuracy follow from the Chain Ladder's characteristics. In practice, assuming the ccODP model is applied, some of these weaknesses can be easily mitigated.
For example, a low value for $X_{40,1}$ will lead the ccODP to estimate low losses for all of AQ40. Hence, in this paper, if the ccODP coefficient for AQ40, $ln(A_{40})$, is below the average of all AQ coefficients, it is taken as an average of the previous 3 quarters, as such:
\begin{equation}
\text{If } ln(A_{40}) < \frac{\sum_{i = 1}^{40}ln(A_i)}{40} \text{, then }ln(A_{40}) \leftarrow \frac{ln(A_{37}) + ln(A_{38}) + ln(A_{39})}{3}.
\end{equation}

\subsection{Mixture Density Network (MDN)}

In this paper, we use Mixture Density Networks (MDNs) to perform probabilistic forecasting of outstanding claims. 

\subsubsection{Distribution of the incremental claims}

Incremental claims $X_{i,j}$ are assumed to follow a mixture Gaussian distribution
\begin{equation}
f_{\hat{X}_{i,j}}(x) = \sum_{k = 1}^{K}\alpha_{i,j,k} \phi({x | \mu_{i,j,k},{\sigma_{i,j,k}} } )\label{eq:MDN}
\end{equation}
With that distributional assumption, the output layer of the MDN estimates the parameters of the mixture distribution, $(\boldsymbol{\alpha}, \boldsymbol{\mu}, \boldsymbol{\sigma}$), which are used to form a mixture Gaussian density. A Negative Log Likelihood (NLL) loss function 
\begin{equation}
NLLLoss(\textbf{X}, \hat{\textbf{X}}|\textbf{w}) = -\frac{1}{|\textbf{X}|}\sum_{i,j: X_{i,j} \in Train} ln(f_{\hat{X}_{i,j}}(X_{i,j} | \textbf{w}))\label{eq:NLL}
\end{equation}
is used to train the MDN, where $\textbf{X}$ is the set of cells $X_{i,j}$ in the training set, $|\textbf{X}|$ is the cardinal of $\textbf{X}$, $\hat{\textbf{X}}$ is the set of predicted distributions of $X_{i,j}$ and $\textbf{w}$ is the set of weights in the MDN. 

The MDN isn't structurally restricted to fitting mixture Gaussians to the response; it can estimate the parameters of any desired distribution so long as the loss function is specified accordingly. In this paper, only the mixture Gaussian framework was considered, with the output layer estimating the $(\boldsymbol{\alpha},\boldsymbol{\mu} ,\boldsymbol{\sigma}  )$ parameters.

As \citet*{Bi1994} notes, the mixture Gaussian distribution, given a sufficient number of components and hidden layers, is capable of approximating any desired distribution within a desired accuracy. Mixture densities with more components will certainly be more flexible, however, practical obstacles such as over-parametrisation and data insufficiency will limit the range of distributions fit by the MDN. 

Alternatively, while maintaining a mixture Gaussian output layer, a mixture \emph{Log-Gaussian} can also be fit to $X_{i,j}$ by fitting a mixture Gaussian to $ln(X_{i,j})$---This holds by the definition of a mixture random variable; see Appendix \ref{app:log} for details). This distribution helps to address the practical limitations to the flexibility of the mixture Gaussian by providing a positive, heavier-tailed option. Furthermore, taking the log of incremental claims linearises the data, which can make training simpler and more efficient. Both the mixture Gaussian and mixture Log-Gaussian distributions achieved impressive results, analysed in Section \ref{sec:Results}.

\begin{remark} We modelled the log of the data where justified (as is often the case in actuarial applications), leading to a mixture Log-Gaussian, but alternative transforms could be readily used by the modeller if needed. \end{remark}

\subsubsection{Structure of the MDN}

\begin{figure}[htb]
\centerline{\includegraphics[width = 8cm]{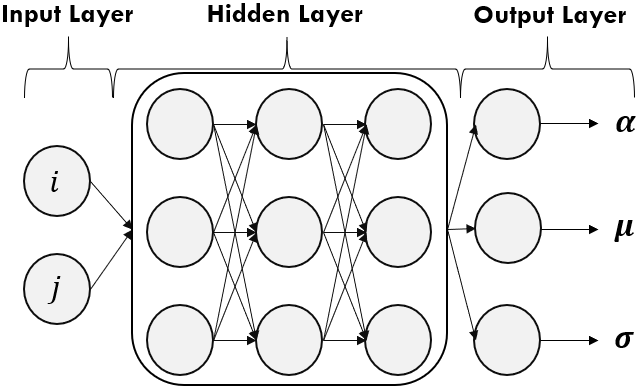}}
\caption{The basic design of the Mixture Density Network (MDN). The inputs $(i,j)$ are the accident and development quarters respectively. The outputs are the parameters of the mixture Gaussian distribution, $\alpha, \mu, \sigma$}
\label{fig:MDN}
\end{figure}

Figure \ref{fig:MDN} provides a basic visualisation of the MDN's design. Mixture Density Networks differ from other neural networks due to their output layer, which estimates a mixture distribution, commonly mixture Gaussian, to the response variable. For a detailed overview of neural networks, their design, mechanism and terminology, see, for instance, \citet*{Ri2018}. In the case of aggregate loss triangles that we consider, the input variables of the MDN are $i$ and $j$, the accident and development periods, respectively. These variables are passed through a fully connected hidden layer, which consists of hidden layers, each layer consisting of neurons. Each neuron in a hidden layer takes a weighted sum of the previous layer's output, before passing it through an activation function. The final hidden layer's output is then passed to the output layer, which produces the desired distribution parameters.

Let $w_{a,c}^l$ be the weight parameter connecting the $a^{th}$ neuron in the $l^{th}$ layer to the $c^{th}$ neuron in the $(l+1)^{th}$ layer. Define $b_{a}^{l-1}$ as the bias term added to the $a^{th}$ neuron in the $l^{th}$ layer. Weighted sums of the inputs, $(i,j)$, are passed into the first hidden layer, before an activation function $g$ yields such output for the $p^{th}$ node in that layer:
\begin{equation}
\textbf{z}_{i,j,p}^1 = g(i \times w_{1,p}^0 + j\times w_{2,p}^0 + b_{p}^0).
\end{equation}
We assume $L$ hidden layers in the MDN, with $D$ nodes in each layer. Each node in successive hidden layers take a weighted sum of the output from nodes in the previous layer, such that $\textbf{z}_{i,j,p}^L$, the $p^{th}$ node in the final hidden layer, is calculated as
\begin{equation}
\textbf{z}_{i,j,p}^L = g(\sum_{d = 1}^{D}w_{d,p}^{L-1}\textbf{z}_{i,j,d}^{L-1} + b_{p}^{L-1}).  
\end{equation}
The output layer is split into three sections, each with $K$ nodes, $K$ being the number of components in the Mixture Density. Let's call these sections the alpha, mu and sigma sections, respectively. Similarly to the hidden layers, each node in the output layer takes a weighted sum of the output of all nodes in the last hidden layer, Layer $L$. The weighted sums are then passed through different activation functions for each section to yield the final output of the MDN, $(\boldsymbol{\alpha},\boldsymbol{\mu} ,\boldsymbol{\sigma}  )$. Specifically:
\begin{description}
\item[\textbf{Alpha:}] The output of the $k^{th}$ nodes of the alpha section
\begin{equation}
\textbf{z}_{i,j,k}^\alpha = \sum_{d = 1}^{D}w_{d,k}^{L, \alpha}\textbf{z}_{i,j,k}^{L} + b_{k}^{L, \alpha} \text{, for } k = 1, 2, ..K
\label{eq:alpha1}
\end{equation}
leads to
 \begin{equation}
 \alpha_{i,j,k}    = \frac{e^{\textbf{z}_{i,j,k}^\alpha}}{\sum_{k = 1}^{K}e^{\textbf{z}_{i,j,k}^\alpha}}  .   
 \label{eq:alpha2}
 \end{equation}
Note that the output $\textbf{z}_{i,j,k}^\alpha$ was passed through a Softmax activation function, which ensures that $ \sum_{k = 1}^{K} \alpha_{i,j,k} = 1 $.
\item[\textbf{Mu:}] Similarly,
\begin{equation}
\textbf{z}_{i,j,k}^\mu = \sum_{d = 1}^{D}w_{d,k}^{L, \mu}\textbf{z}_{i,j,k}^{L} + b_{k}^{L, \mu} \text{, for } k = 1, 2, ..K
\label{eq:mu1}
\end{equation}
leads to
\begin{equation}
    \mu_{i,j,k} = \textbf{z}_{i,j,k}^\mu     
    \label{eq:mu2}
\end{equation}
as there are no constraints on the mu layer which would require an activation function. \citet*{Bi1994} notes that such a design represents an `un-informative prior' on $\mu$, which befits the lack of constraints on the mean.
\item[\textbf{Sigma:}] Finally, the sigma output
\begin{equation}
\textbf{z}_{i,j,k}^\sigma = \sum_{d = 1}^{D}w_{d,k}^{L, \sigma}\textbf{z}_{i,j,k}^{L} + b_{k}^{L, \sigma} \text{, for } k = 1, 2, ..K
\label{eq:sigma1}
\end{equation}
is passed through an exponential function,
\begin{equation}
  {\sigma_{i,j,k}} =   e^{\textbf{z}_{i,j,k}^\sigma},
  \label{eq:sigma2}
\end{equation}
which ensures the standard deviation is always positive \citep*{HjNa2000}.
\end{description}
Here, $w_{d,k}^{L,\alpha}, w_{d,k}^{L,\mu}, w_{d,k}^{L,\sigma}$ are the weights connecting the output of node $d$ in layer $L$ to node $k$ in the alpha, mu and sigma layers, respectively. Thus, for each input cell $(i,j)$, a unique combination of parameters, $$(\alpha_{ i, j, 1}, \alpha_{ i, j, 2}.....\alpha_{ i, j, K}, \mu_{ i, j, 1}, \mu_{ i, j, 2}.....\mu_{ i, j, K}, \sigma_{ i, j, 1}, \sigma_{ i, j, 2}.....\sigma_{ i, j, K} ),$$is produced in the output layer of the MDN, which then generates the probability density for $\hat{X}_{i,j}$ 
\begin{equation}
f_{\hat{X}_{i,j}}(x)  =  \sum_{k = 1}^{K} \alpha_{i,j,k} \phi(x | \mu_{i,j,k}, \sigma_{i,j,k}).\label{eq:MDN2}
\end{equation}

\subsection{Boosting GLM based models with the MDN: ResMDN}

The Mixture Density Network (``MDN'') has greater computational complexity than the standard feedforward neural network, hence its interpretability is even lower. To implement a more interpretable structure, this paper adapts the residual neural network (``ResNet'') design implemented successfully by \citet*{GaRiWu2020}, \citet*{Ga2019} and \citet*{Po2019}, which boosted a GLM model with a neural network - resulting in a more interpretable and stable model. In this paper, this boosting design was adapted to the MDN to create the``ResMDN''. Note that in our ResMDN approach the mean of the resulting model can be interpreted as a boosted version of the GLM backbone, but the other probabilistic distributional properties for the resulting model are inherited from the MDN. This makes the ResMDN approach in principle quite different from the ResNet approach by \citet*{GaRiWu2020}, which focused on the mean.

The ResMDN uses a skip connection, applied in the form of an Embedding Layer, to connect the input layer directly to the output layer. This skip connection allows the MDN to initialise with an approximate GLM fit, subsequently enabling the feedforward module to boost the GLM during training.

In the following, we chose to illustrate the ResMDN by boosting the well known (and used) ccODP model. It is worthwhile to note that---quite naturally---some of the benefits and drawbacks of the ccODP will flow on to the outcomes of the resulting ResMDN model to a certain extent. Indeed, this is illustrated in Section \ref{sec:practical}.

\subsubsection{Distribution of the incremental claims}

 The ResMDN embeds an approximation of the Cross-Classified Over-Dispersed Poisson (ccODP) model (see Section \ref{sec:ccODP} for more detail), which follows the distribution outlined in \eqref{eqn:ccODP}. In neural network terminology, an embedding is the mapping of a discrete or categorical input variable into a numerical vector, which is then fed into the network \citep{Ri2018}.
 Since the output of the ResMDN takes the form of parameters for a mixture Gaussian distribution, the GLM initialisation's density is approximated by a mixture Gaussian, spread evenly over $K$ components. The parameters of this approximation are fed into the ResMDN as embeddings of the GLM. The distribution $ f_{\hat{X}_{i,j}^{ccODP}}(x)$ of $X_{i,j}$ as estimated by the ccODP is approximated by
\begin{equation}
 f_{\hat{X}_{i,j}^{ccODP}}(x) \approx  \sum_{k = 1}^{K}\alpha^{GLM}_{i,j,k} \phi_{i,j,k}(x | {\mu^{GLM}_{i,j,k}, \sigma^{GLM}_{i,j,k}}), \label{eq:MDNODP}
\end{equation}
where
\begin{equation}
\alpha_{i,j,k}^{GLM} = \frac{1}{K}, \quad \mu_{i,j,k}^{GLM} = E[\hat{X}_{i,j}^{ccODP}] = A_iB_j, \quad \sigma_{i,j,k}^{GLM} = \sqrt{Var[\hat{X}_{i,j}^{ccODP}]} = \sqrt{D A_iB_j}.
\end{equation}

\subsubsection{ResMDN structure}
The ResMDN's structure resembles the MDN very closely. The Input Layer of the ResMDN consists of the accident and development periods, $(i,j)$, as well as a unique categorical integer, $c_{i,j} = 40*(i-1) + j$, which allows the ResMDN's embedding layer to identify the specific cell $(i,j)$ and produce the corresponding GLM loss estimate for that cell as output. Hence, each cell $(i,j)$ is assigned a number from $(0,1599)$. The variables $(i,j)$ are passed through an MDN which excludes the activations of the final output layer. An embedding layer takes the categorical input $c_{i,j}$ and produces as output: 
\begin{equation}
\bigg(ln(\boldsymbol{\alpha^{GLM}_{i,j}}), \boldsymbol{\mu^{GLM}_{i,j}}, ln(\boldsymbol{\sigma^{GLM}_{i,j}}) \bigg)
\end{equation}
The outputs from both the fully connected component and embedding layer are added together, before the Softmax and exponential activations are applied to the alpha and sigma additions, respectively. The final output of the ResMDN consists of the mixture Gaussian parameter estimates, $(\boldsymbol{\alpha^{ResMDN}}, \boldsymbol{\mu^{ResMDN}}, \boldsymbol{\sigma^{ResMDN}}  )$. Figure \ref{fig:ResMDN} provides a visualisation of the ResMDN model. The key design features of the ResMDN are outlined below:

\begin{figure}[htb]
\centerline{\includegraphics[width = 8cm]{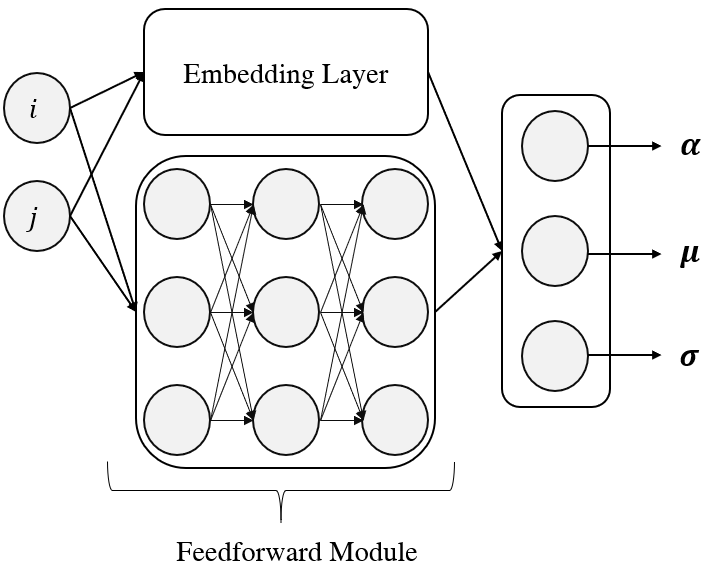}}
\caption{The ResMDN design with a mixture Gaussian output. The embedding layer converts the input to mixture Gaussian parameters approximating the GLM backbone. The feedforward module boosts the GLM initialisation during training.}
\label{fig:ResMDN}
\end{figure}

\begin{itemize}
    \item \textbf{Fully Connected Module:} Let $\textbf{z}_{i,j,k}^\alpha, \textbf{z}_{i,j,k}^\mu, \textbf{z}_{i,j,k}^\sigma$ be as described in \eqref{eq:alpha1}, \eqref{eq:mu1} and \eqref{eq:sigma1}, that is, the MDN's output before the output layer activations are applied. The fully connected module of the ResMDN performs the function
\begin{equation}
(i,j) \mapsto \{(\textbf{z}_{i,j,k}^\alpha, \textbf{z}_{i,j,k}^\mu, \textbf{z}_{i,j,k}^\sigma), k = 1,2,..K \},
\label{eq:FullyConnected}
\end{equation}
\item \textbf{Embedding Layer:} The embedding layer weights are pre-set to provide the mapping 
\begin{equation}
 c_{i,j} \mapsto \bigg(ln(\boldsymbol{\alpha^{GLM}_{i,j}}), \boldsymbol{\mu^{GLM}_{i,j}}, ln(\boldsymbol{\sigma^{GLM}_{i,j}}\bigg).
 \label{eq:Embedding}
\end{equation}
The log of the $\boldsymbol{\alpha}$ and $\boldsymbol{\sigma}$ parameters are produced in the embedding layer, since the Softmax and exponential Activation functions will take the exponent in the output layer nodes. \item \textbf{Addition and Final Activation:} The output from the embedding and fully connected modules are added together element-wise: 
\begin{align}
 \textbf{Addition: }&\bigg(ln(\alpha^{GLM}_{i,j,k}), \mu^{GLM}_{i,j,k}, ln(\sigma^{GLM}_{i,j,k}), \textbf{z}_{i,j,k}^\alpha, \textbf{z}_{i,j,k}^\mu, \textbf{z}_{i,j,k}^\sigma \bigg)  \nonumber \\
 & \mapsto \bigg( ln(\alpha^{GLM}_{i,j,k}) +  \textbf{z}_{i,j,k}^\alpha, \mu^{GLM}_{i,j,k} +  \textbf{z}_{i,j,k}^\mu, ln(\sigma^{GLM}_{i,j,k}) + \textbf{z}_{i,j,k}^\sigma\bigg)  \nonumber \\
 & \quad= \bigg(  ln(\alpha^{GLM}_{i,j,k}) +  \textbf{z}_{i,j,k}^\alpha,   \mu^{ResMDN}_{i,j,k}, ln(\sigma^{ResMDN}_{i,j,k}) \bigg),
 \label{eq:ResMDNAddition}
\end{align}
before the Softmax and exponential activations are applied to the alpha and sigma layers, respectively:
\begin{align}
\textbf{Final Activations: }&\bigg(ln(\alpha^{GLM}_{i,j,k}) +  \textbf{z}_{i,j,k}^\alpha, \mu^{ResMDN}_{i,j,k}, ln(\sigma^{ResMDN}_{i,j,k})\bigg)  \nonumber \\
& \mapsto \left(\frac{{\alpha^{GLM}_{i,j,k} e^{\textbf{z}_{i,j,k}^\alpha}}}{\sum_{k = 1}^{K}\alpha^{GLM}_{i,j,k} e^{\textbf{z}_{i,j,k}^\alpha}}, \mu^{ResMDN}_{i,j,k}, e^{ln(\sigma^{ResMDN}_{i,j,k})} \right) \nonumber \\
& \quad = \left(\alpha^{ResMDN}_{i,j,k}, \mu^{ResMDN}_{i,j,k}, \sigma^{ResMDN}_{i,j,k} \right).
\label{eq:ResMDNFinal}
\end{align}
Hence, the boosted mixture Gaussian parameters are produced in the Output Layer. 

\item \textbf{Initialisation:} The activations, $\textbf{z}_{i,j,k}^\alpha, \textbf{z}_{i,j,k}^\mu, \textbf{z}_{i,j,k}^\sigma$, are generated using the parameters, $(\textbf{w}_L, \textbf{b}_L)$, defined in \eqref{eq:alpha1}--\eqref{eq:sigma2}. These parameters, representing the weights in the final hidden layer, are initialised at 0, such that
$$\textbf{z}_{i,j,k}^\alpha, \textbf{z}_{i,j,k}^\mu, \textbf{z}_{i,j,k}^\sigma = 0, \alpha_{i,j,k}^{ResMDN}  = \alpha_{i,j,k}^{GLM} , \mu_{i,j,k}^{ResMDN}  = \mu_{i,j,k}^{GLM} , \sigma_{i,j,k}^{ResMDN}  = \sigma_{i,j,k}^{GLM}, $$
hence producing the GLM approximation in the Output Layer at the initialisation of the ResMDN. This initialisation follows the methodology of \citet*{GaRiWu2020} closely.

\end{itemize}

 During training, the embedding layer maintains constant output, while the fully connected module adjusts its weights to capture non-linearities which the GLM has missed. The ResMDN's overall function at the termination of training is 
\begin{align}
(i,j, c_{i,j}) &\mapsto \left(\frac{{\alpha^{GLM}_{i,j,k} e^{\textbf{z}_{i,j,k}^\alpha}}}{\sum_{k = 1}^{K}\alpha^{GLM}_{i,j,k} e^{\textbf{z}_{i,j,k}^\alpha}}, \mu^{GLM}_{i,j,k} + \textbf{z}_{i,j,k}^\mu, {\sigma^{GLM}_{i,j,k} e^{\textbf{z}_{i,j,k}^\sigma}} \right)  \nonumber \\
&\quad= \left(\alpha^{ResMDN}_{i,j,k}, \mu^{ResMDN}_{i,j,k}, \sigma^{ResMDN}_{i,j,k} \right)\text{, for } k = 1,2,...,K.
\label{eq:ResMDNFinalEqn}
\end{align}
The NN boosting terms are relatively easy to analyse in relation to the GLM fit, especially the mean and volatility terms. Furthermore, the black box neural network modelling is only applied to the residuals, meaning the lack of interpretability is restricted to that domain only. Hence the ResMDN improves the interpretability of the model compared to the MDN. 
\section{Model Development}
\label{sec:ModelDev}

\subsection{Model selection using the rolling origin method for training, validating, and testing}
The accuracy of the neural network depends heavily on its hyper-parameters, such as the number of hidden layers, number of neurons, weight regularisation penalty, etc. Therefore, to assume that one model design will work well in all environments will lead to sub-optimal performance. Hence, a training/testing split is required to assess different model designs and choose the best one found. This paper partitions the loss triangle using the rolling origin method, which performs the training and testing split in multiple stages, each one progressively shifting the testing set forward in time (see \citet*{Ta2000,BeBe2012,BaRi2020} for details). The total test error of the model is a weighted average of the test error in each stage. This methodology allows for a systematic hyper-parameter fine-tuning algorithm to be implemented; see Section \ref{sec:algorithm}.

The loss triangle has the characteristics of a time series, with the incremental claims generally decaying over successive development periods. Where the objective of modelling is to improve interpolation accuracy, randomly splitting the data into training, validation and testing sets is common and sufficient. With loss triangles, however, the objective is \emph{extrapolation}, hence the testing set needs to focus on assessing the model's \emph{projection accuracy}. This is done by assigning (a chosen number of) the \emph{latest} calendar periods of the triangle to the \emph{testing} set and the \emph{earliest} to \emph{training}. Similarly, the Validation set is chosen to be the latest calendar periods which aren't assigned for testing. That way, when combined with Early Stopping (see Section \ref{sec:training}), the MDN stops training when short term projection accuracy is maximised.

\begin{figure}[htb]
\centering
\begin{subfigure}{.5\textwidth}
  \centering
  \includegraphics[width=.9\linewidth]{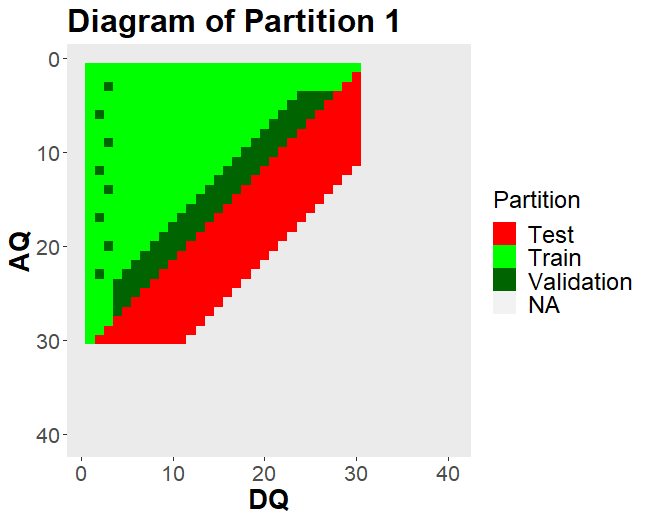}
  \caption{Partition 1: Assesses projection accuracy}
  \label{fig:sub1}
\end{subfigure}%
\begin{subfigure}{.5\textwidth}
  \centering
  \includegraphics[width=.9\linewidth]{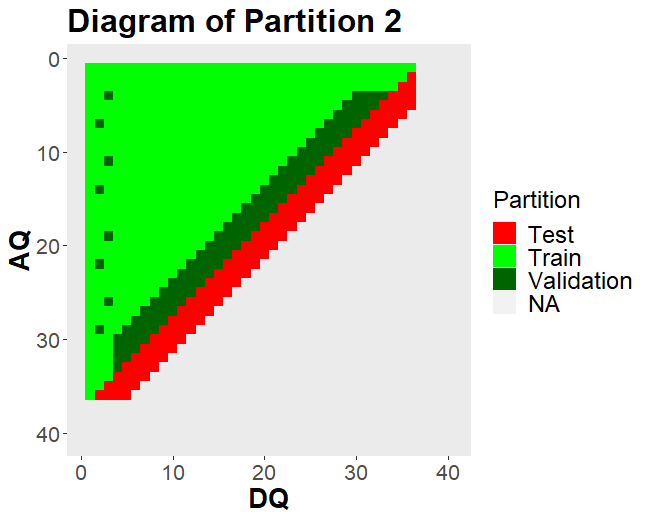}
  \caption{Partition 2: Assesses trend fitting}
  \label{fig:sub2}
\end{subfigure}
\caption{The 2-stage partition of the triangle into training, validation and testing sets. The first partition focuses on assessing projection accuracy, while the second assesses the model's ability to fit more recent trends in the data.}
\label{fig:rollingorigin}
\end{figure}

Hence, it is important to \emph{sequentially} split the data into training, validation and testing sets to more effectively assess the model's accuracy when extrapolating. In our illustrative example of a 40$\times$40 triangle, the rolling origin validation method was used in two partitions:
\begin{itemize}
\item In the first partition, the data is assumed to comprise a 30$\times$30 triangle, which leaves the latest 10 calendar periods for the testing set. This partition focuses on assessing the model’s long term forecasting accuracy.
\item The second partition works with a 36$\times$36 triangle, leaving 4 calendar periods for testing. Building on the first partition, later calendar periods are included in training, which helps to assess the model's ability to capture the more recent and more holistic trends present in the triangle.
\end{itemize}
For all partitions, the validation set included the 4 latest non-testing calendar periods, excluding the first 3 accident and development periods. This exclusion was done to provide the MDN more training data for the latest accident and development periods. Instead, the DQ2 and DQ3 validation points are taken evenly from earlier AQs, an arbitrary but simplistic approach. Figure \ref{fig:rollingorigin} visualises the data partitions. 

A potential downside of the rolling origin method is that the training data does not include points from the latest calendar quarters. Therefore, in situations where we expect losses to possess substantially different characteristics in the later periods, this method might not be able to effectively capture the change in trends. In such situations, we implement an adjusted data partitioning methodology to allow more training data in those periods (see Section \ref{app:partition}).

\subsection{Network hyper-parameter selection algorithm }
\label{sec:algorithm}
There is a vast array of literature surrounding neural network model search and selection, and that no set algorithm succeeds the rest. %We refer to \cite{Ab2019} for details.
The MDN's architecture was selected using an algorithm that successively optimised one hyper-parameter at a time. The number of components in the density is increased so long as the test error decreases, which allows the algorithm to consider fitting densities of infinite flexibility. The loss triangle size restricts the parametrisation of the mixture density, hence this algorithm aims to find the optimal distributional flexibility allowed by the given data. 
Before the algorithm is run, certain aspects of the MDN's design are set constant:
\begin{itemize}
\item The sigmoid activation function is used for all hidden layers;
\item The number of neurons is equal for all hidden layers.
\end{itemize}

\noindent With this modelling framework, two additional considerations improved the MDN's performance:
\begin{itemize}
    \item Mixture Gaussian or mixture Log-Gaussian: Fitting a mixture Log-Gaussian linearises the data (in our methodology), which provided smoother results in some instances. When used, the mixture Log-Gaussian gave noticeably improved results over the mixture Gaussian in the upper triangle.
    \item Mean Squared Error term to the loss function: A Negative Log-Likelihood loss function sometimes failed to capture sharp points in the data, preferring to allocate volatility to these points. This issue was solved by adding a Mean Squared Error (``MSE'') term to the loss, which encouraged the MDN to provide more accurate central estimates. 
\end{itemize}

\noindent The hyper-parameters chosen to fine-tune are:
\begin{enumerate}
\item $\lambda_w$, the L2 weight penalty. The values $[0,0.0001,0.001,0.01,0.1]$ were tested.
\item $\lambda_\sigma$, the L2 sigma activity penalty. When central estimates are inaccurate, the MDN may increase volatility estimates unreasonably to reduce the Negative Log Likelihood Loss \eqref{eq:NLL} ( \citet[Ch6.2]{GoBeCo2017}), hence a penalty was applied to the sigma output. The values $[0,0.0001,0.001,0.01,0.1]$ were tested.
\item $p$, the dropout rate (See \citet*{SrHiKrSu2014} for details). The values $[0,0.1,0.2]$ were tested. 
\item $n$, the number of neurons in each hidden layer. The values $[20,40,60,80,100]$ were tested.
\item $h$, the number of hidden layers. The values $[1,2,3,4]$ were tested.
\item $K$, the number of components in the mixture density.
\end{enumerate}

\noindent The training loss becomes
\begin{align}
Loss(\textbf{X}, \hat{\textbf{X}}|\textbf{w}, \lambda_w, \lambda_{\sigma} ) &= - \frac{1}{|\textbf{X}|}\sum_{i,j: X_{i,j} \in \textbf{X}} ln(f_{\hat{X}_{i,j}}(X_{i,j} | \textbf{w})) \nonumber \\
&+ \lambda_w\textbf{w}\cdot\textbf{w} +  \lambda_{\sigma}\sum_{i,j: X_{i,j} \in \textbf{X}}\sum_{k = 1}^{K}\sigma_{i,j,k}^2 
\label{eq:trLoss}
\end{align}

Denote $\boldsymbol{\theta} = \{ \lambda_w, \lambda_\sigma, p, n, h, K \}$ as the set of hyper-parameters to fine-tune. The hyper-parameter selection algorithm was conducted as such: 
\begin{enumerate}
\item Start with $\boldsymbol{\theta}^{initial} = \{ 0,0,0,n^{initial}, h^{initial}, K^{initial}  \}$, a set of initial hyper-parameters deemed suitable through judgement. Setting $\boldsymbol{\theta}^{initial} = \{ 0,0,0,60, 2, 2  \}$ worked well in this paper, as allowing the algorithm to explore unregularised models vastly improved the fit in some instances.
\item Using $\boldsymbol{\theta}^{initial}$ and keeping all other hyper-parameters fixed, use \textbf{Grid Search} to test all desired values of $\lambda_w$, the weight penalty coefficient. Select the coefficient with the lowest test error, $\hat{\lambda}_w$, and update $\boldsymbol{\theta}^1 = \{ \hat{\lambda}_w,0,0,n^{initial}, h^{initial}, K^{initial}  \}$
\item Using $\boldsymbol{\theta}^1$ and keeping all other hyper-parameters fixed, use \textbf{Grid Search} to test all desired values of $\lambda_\sigma$, the sigma activity penalty coefficient. Select the coefficient with the lowest test error, $\hat{\lambda}_\sigma$, and update $\boldsymbol{\theta}^2 = \{ \hat{\lambda}_w,\hat{\lambda}_\sigma,0,n^{initial}, h^{initial}, K^{initial}  \}$
\item  Using $\boldsymbol{\theta}^2$ and keeping all other hyper-parameters fixed, use \textbf{Grid Search} to test all desired values of $p$, the dropout rate. Select the rate with the lowest test error, $\hat{p}$, and update $\boldsymbol{\theta}^3 = \{ \hat{\lambda}_w,\hat{\lambda}_\sigma,\hat{p},n^{initial},\linebreak[0] h^{initial}, K^{initial}  \}$
\item Using $\boldsymbol{\theta}^3$ and keeping all other hyper-parameters fixed, use \textbf{Grid Search} to test all desired values of $h$, the number of hidden layers. Select the number with the lowest test error, $\hat{h}$, and update $\boldsymbol{\theta}^4 = \{ \hat{\lambda}_w,\hat{\lambda}_\sigma,\hat{p},n^{initial}, \hat{h}, K^{initial}  \}$
\item Using $\boldsymbol{\theta}^4$ and keeping all other hyper-parameters fixed, fine-tune the number of neurons and components. This process tests an increasing number of components until the test error ceases to improve. Let $n_K$ be the number of neurons which minimises the test error for a $K$-component model (among the values tested). Let $E_{n_K, K}$ be the test error of a model with $n_K$ neurons and $K$ components (with the other hyper-parameters as in $\boldsymbol{\theta}^4$). Starting at $K = 1$, increment $K$ until $E_{n_K, K} < E_{n_{K+1}, K+1}$. At this final increment, set $\hat{K} = K$ and $\hat{n} = n_{K}$. Update the hyper-parameters, $\boldsymbol{\theta}^5 = \{ \hat{\lambda}_w,\hat{\lambda}_\sigma,\hat{p},\hat{n}, \hat{h}, \hat{K}  \}$
\end{enumerate}
Following the algorithm, select $\boldsymbol{\theta}^5$ as the final set of hyper-parameters and run the final model (see Section \ref{sec:training}.

\subsection{Data}
\label{sec:Data}
This paper tests the MDN's performance on both simulated and real data. Using simulated data allows the practitioner to simulate any desired trend, providing a controlled environment where the MDN can be directly assessed on its ability to capture these trends embedded in the data \citep*{AvTaWaWo2020}. A downside of simulating claims, as noted by \citet*{MuRyRe2011} is that complex interactions in real data may not be captured by the simulator, hindering model development. Fitting a model on unrealistic data will reduce its validity. However, we mitigate this by applying the MDN on real data (AUSI environment; see Section \ref{sec:AUSI}) as well as the default SynthETIC dataset from \citet*{AvTaWaWo2020}, which can mimic data in a wide range of realistic situations. 

Note that while the AUSI Dataset was partitioned randomly into 10 triangles (via subdivision), we simulated 200 triangles (from four separate realistic scenarios) with SynthETIC, meaning that the MDN was fit on 210 triangles in total.

\subsubsection{Simulated environments from SynthETIC}

 Thanks to the flexibility offered by SynthETIC  \citep*[see][for a description of the R simulation package]{AvTaWaWo2020}, four different claim environments were simulated, of various features and complexities. Simulating different environments was done to test the MDN's versatility and ability to capture complex trends and produce accurate forecasts in a variety of controlled, challenging environments. As a loss triangle is a collection of random variables, it is important to run the MDN on a large sample of triangles to gain a better understanding of its accuracy and also test its ability to provide consistent results. Hence, for each simulated environment, 50 independent triangles were simulated (of size $40\times 40$), leading to 200 triangles. Namely, the scenarios, or environments, were:
\begin{enumerate}
\item \textbf{Environment 1 - Simple, short tail claims: } This environment simulates short tail claims which are homogeneous in composition for all accident quarters. The reporting and settlement delays have been approximately calibrated to show similar characteristics to the simulator developed by \citet*{GaWu2018}. Figure \ref{fig:D1} plots the incremental claims for Environment 1. A spike in claim payment in DQ2 complicates this dataset, but such a feature is not uncommon in practice. This environment was a preliminary test to the feasibility of MDNs in modelling 40x40 triangles and producing reasonable results. 

\begin{figure}[htb]
\centerline{\includegraphics[width = 10cm]{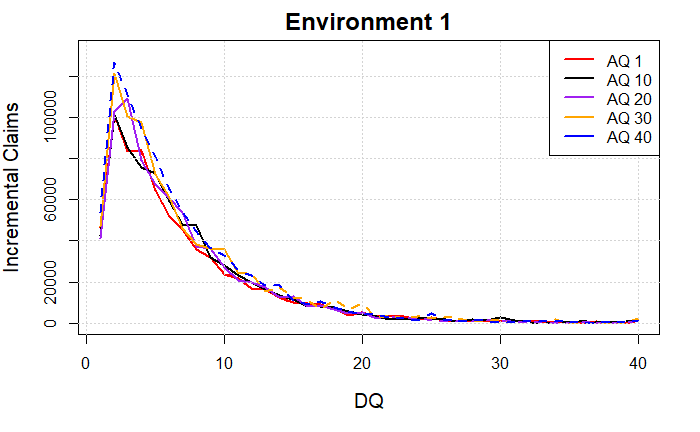}}
\caption{A plot of the incremental claims of environment 1, for selected accident quarters. Solid lines represents data in the upper triangle, while dashed lines represent data in the lower triangle   }
\label{fig:D1}
\end{figure}
\item \textbf{Environment 2 - Increase in claim processing speed:} A gradual shift from long tail to short tail claims along accident quarters is simulated, that is, an increase in claims processing speed. Initially, there are more long tail claims, however, the proportion of these claims decreases, while the proportion of short claims increases. From the incremental claim plot shown below, later AQs see higher losses early on, due to the increasing proportion of short tail claims. Figure \ref{fig:D2} plot the incremental claims of environment 2. The main question to be answered in testing this dataset is, given the systematic volatility in the claims data, can the MDN accurately distinguish between systematic and unsystematic volatility and capture the distribution of data points accurately? That is, will the MDN learn that claims are getting shorter, or will it attribute the trend to noise?\begin{figure}[htb]
\centerline{\includegraphics[width = 10cm]{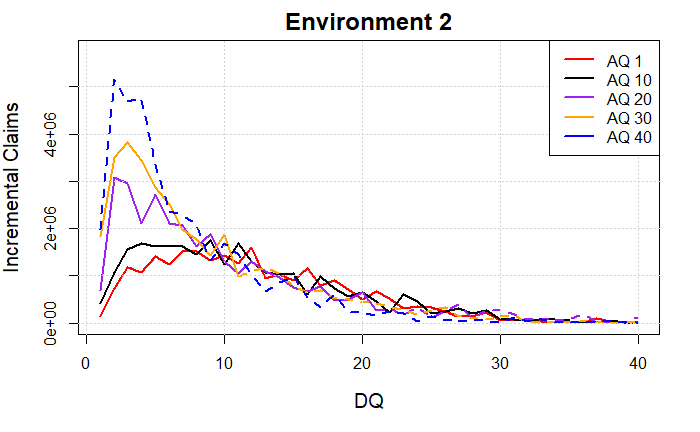}}
\caption{A plot of the incremental claims of environment 2, for selected accident quarters. Solid lines represents data in the upper triangle, while dashed lines represent data in the lower triangle   }
\label{fig:D2}
\end{figure}
\item \textbf{Environment 3 - Inflation shock:} Superimposed inflation is changed instantly from 0\% to 8\% per annum, starting at AQ30. The 8\% inflation remains constant in the lower triangle. This environment tests the ability of the MDN to recognise changes in calendar effects and adapt projections accordingly.

Only the last 10 calendar quarters in the upper triangle contain information regarding the inflation shock, which increases the difficulty for the MDN. A further complication is that the rolling origin partition contains little training points featuring the change in inflation, hence this environment assesses its ability to capture recent trends. Figure \ref{fig:D3} plots the incremental claims.
\begin{figure}[htb]
\centerline{\includegraphics[width = 10cm]{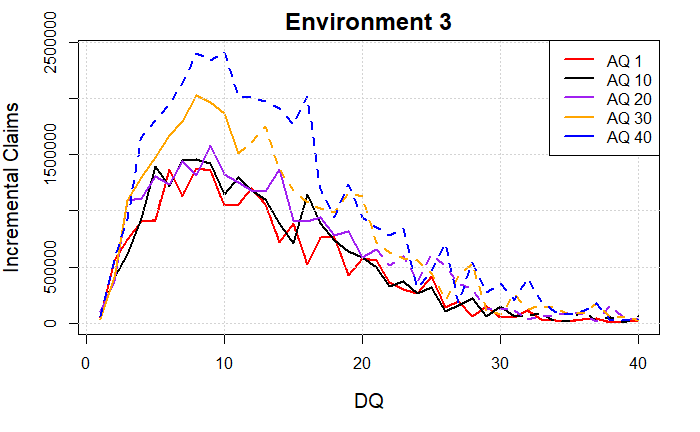}}
\caption{A plot of the incremental claims of environment 3, for selected accident quarters. Solid lines represents data in the upper triangle, while dashed lines represent data in the lower triangle   }
\label{fig:D3}
\end{figure}
\item \textbf{Environment 4 - High Systematic Complexity:} This environment is the default triangle generated by the SynthETIC simulator, which was designed to mimic features seen in real data \citep*[see][for details]{AvTaWaWo2020}. Complex dependencies exist between claim size, reporting and settlement delay, and superimposed inflation. Settlement delay, which depends on claim size, declines over the first 20 AQs. Superimposed inflation is up to 30\%, but declines for larger claims. A legislative change at AQ20 causes small claims to face a reduction in size and settlement speed until AQ30. The general trend can be summarised by slow development, high volatility and high superimposed inflation. The volatility is primarily caused by the low claim frequency and highly volatile severity. Hence, it is normal for claims in one AQ to follow a completely different pattern (reporting, settlement, volume, development pattern) than claims in the adjacent AQ. Figure \ref{fig:D4} plots the incremental claims. This environment assesses the MDN's ability to produce accurate forecasts in a volatile environment.
\begin{figure}[htb]
\centerline{\includegraphics[width = 10cm]{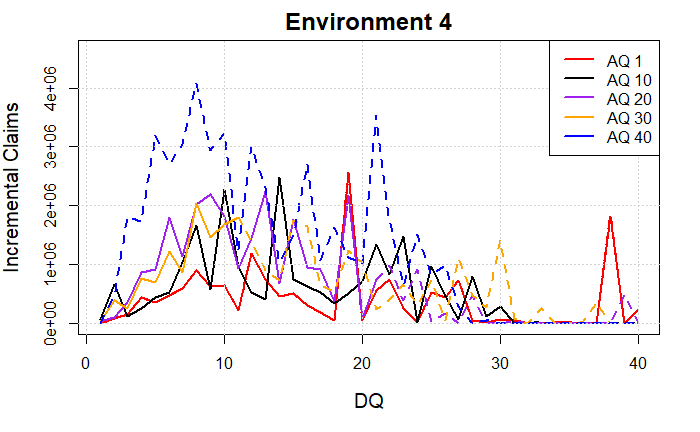}}
\caption{A plot of the incremental claims of environment 4, for selected accident quarters. Solid lines represents data in the upper triangle, while dashed lines represent data in the lower triangle   }
\label{fig:D4}
\end{figure}
\end{enumerate}

\subsubsection{Real Dataset - AUSI Auto Bodily Injury}
\label{sec:AUSI}
\begin{figure}[htb]
\centerline{\includegraphics[width = 10cm]{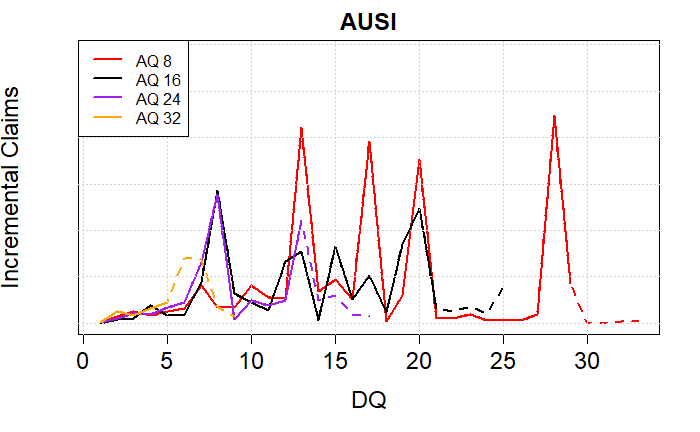}}
\caption{A plot of the incremental claims of the AUSI environment, for selected accident quarters. Solid lines represents data in the upper triangle, while dashed lines represent data in the lower triangle  }
\label{fig:AUSI}
\end{figure}

We apply the MDN to a dataset obtained through a collaborative project between Allianz, University of New South Wales (UNSW), Suncorp and Insurance Australia Group (IAG). This forms the AUSI acronym, which we will use to refer to this dataset. The Auto Bodily Injury line of business is used, which features slow claim development and high volatility. The AUSI dataset consists of transactional data for individual claims, which we aggregated into quarterly triangles. We use quarterly data from January 2005 to December 2014, which provides a 36$\times$36 upper triangle and a 4 quarter forecasting period, which will be used to compare the MDN's results to the ccODP. Using the individual claims data, each claim was randomly allocated to one of ten triangles, meaning that ten aggregate triangles of roughly equal size were created using this dataset. As mentioned earlier, running the MDN on multiple triangles better assessed the model's consistency. Figure \ref{fig:AUSI} plots the incremental claims for this dataset. This environment aims to assess the MDN's ability to provide accurate forecasts for real data. 

\section{Model Training and Evaluation }
\label{sec:training}
The SynthETIC Simulator produces individual claims, which are aggregated into a 40x40 triangle. It is a common procedure in neural network modelling to standardise the input variables, in order to stabilise the training. Through early experimentation, normalising the response as well ($X_{i,j}$) was crucial in achieving convergence during training.

For each hyperparameter combination tested, $\boldsymbol{\theta} = \{ \lambda_w, \lambda_\sigma, p, n, h, K \}$, an MDN with such hyperparameters is trained on the training set of each partition, then projected on to the testing set. The loss function we selected is the Negative Log-Likelihood, with weight and sigma activation penalties applied during training:
\begin{equation}
Loss(\textbf{X}, \hat{\textbf{X}}|\textbf{w}, \lambda_w, \lambda_{\sigma} ) = - \frac{1}{|\textbf{X}|}\sum_{i,j: X_{i,j} \in \textbf{X}} ln(f_{\hat{X_{i,j}}}(X_{i,j} | \textbf{w}))  
+ \lambda_w\textbf{w}\cdot\textbf{w} +  \lambda_{\sigma}\sum_{i,j: X_{i,j} \in \textbf{X}}\sum_{k = 1}^{K}\sigma_{i,j,k}^2. \label{eq:train}
\end{equation}
Via experimentation, the Adam optimiser \citep{KiBa2014}, with a learning rate of 0.001 provided the most stable training compared with other optimisers like RMSProp and Stochastic Gradient Descent. To further minimise over-fitting, Early Stopping was used to stop training as soon as the validation loss was minimised \citep{GoBeCo2017}. The validation loss rarely decreased steadily, hence training was only stopped when it did not hit new lows in the last 1000 epochs. This is referred to as the patience measure in the Keras interface; a lower patience than 1000 would sometimes prematurely stop training. Training would usually last for several thousand epochs, with higher dropout rates and larger networks often requiring up to 10-15 thousand iterations. A 10000 epoch limit was set when running the hyper-parameter optimisation algorithm, in order to increase efficiency.

\subsection{Test error}
Denote $\boldsymbol{\theta}$ as the hyper-parameter values of the MDN being run. In addition, let $f_{\hat{X}_{i,j}} ( x| \textbf{w}, \theta)$ be the density of $\hat{X}_{i,j}$ projected by an MDN with hyper-parameters $\boldsymbol{\theta}$ and weights $\textbf{w}$. Let $T1$ and $T2$ be the set of cells $(i,j)$ in the testing set of the first and second partitions, respectively. A separate MDN is trained $T$ times in each partition; let $\textbf{w}_{p,t}$ be the weights of the $t^{th}$ model trained on the $p^{th}$ partition. The test error of the MDN with hyper-parameters $\boldsymbol{\theta}$ is calculated from \eqref{eq:test1} - \eqref{eq:test3}.
\begin{equation}
 \text{TestError}(\boldsymbol{\theta}, \text{Partition } 1) =  -\frac{1}{T|T1|}\sum_{t = 1}^{T}\sum_{i,j: (i,j) \in T1} ln(f_{\hat{X}_{i,j}} ( X_{i,j}| \textbf{w}_{1,t},\boldsymbol{\theta}))
 \label{eq:test1}
\end{equation}
\begin{equation}
 \text{TestError}(\boldsymbol{\theta}, \text{Partition } 2) =  -\frac{1}{T|T2|}\sum_{t = 1}^{T}\sum_{i,j: (i,j) \in T2} ln(f_{\hat{X}_{i,j}} ( X_{i,j}| \textbf{w}_{2,t}, \boldsymbol{\theta}))
  \label{eq:test2}
\end{equation}

\begin{equation}
  \text{TestError}(\boldsymbol{\theta})  =   \frac{ |T1|*\text{TestError}(\boldsymbol{\theta}, \text{Partition } 1)   +   |T2|*\text{TestError}(\boldsymbol{\theta}, \text{Partition } 2)}{|T1| + |T2|}   
   \label{eq:test3}
\end{equation}
Hence, the MDN is trained $2T$ times for each set of hyper-parameters $\theta$, as each run has a different weight initialisation and hence a different fit. Averaging the error of these runs reduces the impact of random weight initialisations on the performance of the hyper-parameter set $\boldsymbol{\theta}$.

\subsection{Projection constraints}
This paper implements a mechanism for directly constraining central estimates of cells in the lower triangle, thereby directly controlling projections made by the MDN. This method allows for the practitioner's judgement to be incorporated if required. The practitioner can place upper and lower bounds on the central estimates of any desired set of cells $(i,j)$ in the lower triangle and penalise the MDN if its central estimates fall outside those boundaries.

Let $\textbf{C}$ be the set of cells $(i,j)$ in the lower triangle, which have had constraints placed on their projections. Let $C_{i,j}^{Lower}$ and $C_{i,j}^{Upper}$ be the lower and upper constraints of the central estimates for cell $(i,j) \in \textbf{C}$. Let $\hat{\mu}_{i,j} = E[\hat{X}_{i,j}]$. The loss function during training follows \eqref{eq:projConst}:
\begin{align}
   NLLLoss(\textbf{X}, \hat{\textbf{X}}|\textbf{w}) &= -\frac{1}{|\textbf{X}|}\sum_{i,j: X_{i,j} \in Train} ln(f_{\hat{X}_{i,j}}(X_{i,j} | \textbf{w})) \\
   &+ Regularisation + \frac{\lambda_C}{|\textbf{C}|}  \sum_{i,j: (i,j) \in \textbf{C}} [max(0,\hat{\mu}_{i,j} - C_{i,j}^{Upper})]^2 + [max(0,C_{i,j}^{Lower} - \hat{\mu}_{i,j})]^2
   \label{eq:projConst}
\end{align}
Where $Regularisation = \lambda_w\textbf{w}\cdot\textbf{w} +  \lambda_{\sigma}\sum_{i,j: X_{i,j} \in \textbf{X}_{Train}}\sum_{k = 1}^{K}\sigma_{i,j,k}^2$ from \eqref{eq:train}, and $\lambda_C$ is a constraint violation penalty coefficient. The constraints apply a square distance penalty to the loss function if the central estimate of constrained cells in the lower triangle violate the constraints. With a sufficiently high penalty coefficient, the MDN's projection will satisfy the constraints specified, providing projections that are more reasonable. The cells in $\textbf{C}$ are randomly split in half between the training and validation sets, as the validation loss should indicate how well the projection constraints have been met in order for Early Stopping to be used effectively. 

\subsection{Fitting the final model}
\begin{figure}[htb]
\centerline{\includegraphics[width = 8cm]{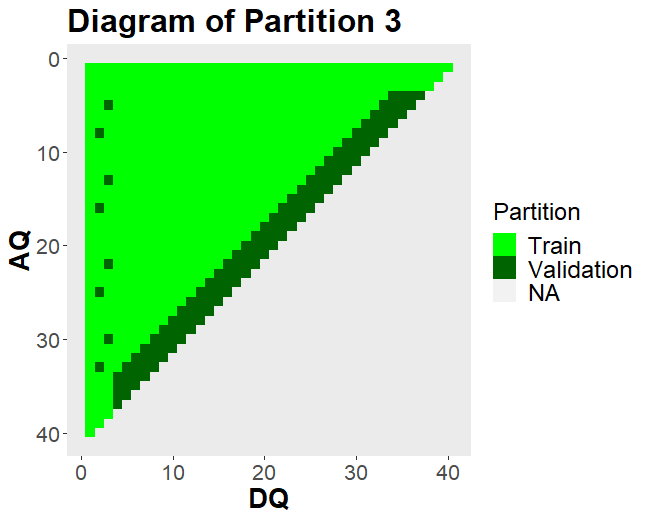}}
\caption{The training/validation partition of the upper triangle. The chosen MDN design is fit on the training data and used to project claims in the lower triangle}
\label{fig:part3}
\end{figure}

Once all desired hyper-parameter combinations are tested, the combination with the lowest test error, $\boldsymbol{\theta}^{min}$ (see Section \ref{sec:algorithm} for details), is set as the model architecture of choice. To produce distributional forecasts of claims in the lower triangle, the chosen MDN is run on the entire upper triangle. Only a training/validation split is needed, since the testing set was only used to compare different hyper-parameters. The training/validation partition of the upper triangle was done sequentially, visualised in Figure \ref{fig:part3}.
An MDN with hyper-parameters $\boldsymbol{\theta}^{min}$ is fit 5 times on the training data of Partition 3, under different weight initialisations. The 5 fitted distributions are ensembled to produce the final forecast. This ensemble of models produces more robust results and reduces the error of bad runs that stop training at a poor local minimum of loss \citep{PeCo1992}. Let $w_z$ be the set of the MDN's final weights in the $z^{th}$ run. The 5 fits are ensembled to produce the distribution of incremental claims shown in \eqref{eq:finaldist}.
\begin{equation}
 f_{\hat{X}_{i,j}}(x) =  \frac{1}{5} \sum_{z = 1}^{5} \sum_{k = 1}^{K} \alpha_{i,j,k}^{w_z} \phi(x | \mu_{i,j,k}^{w_z}, \sigma_{i,j,k}^{w_z}) 
 \label{eq:finaldist}
\end{equation}
\subsection{Model evaluation}\label{sec:metrics}

The final MDN model is fit on the lower triangle and \textbf{compared to the ccODP model}. The results of two main variables were analysed:
\begin{enumerate}
\item Individual cells, $X_{i,j}$
\item Total reserves, $R = \sum_{i,j: i+j > 41}X_{i,j} $
\end{enumerate}
 Capital standards set by APRA and Solvency II require reserve allocations to meet total reserves in the lower triangle with a 75\% and 99.5\% probability of sufficiency, respectively. Hence measuring the accuracy of total reserves is important. However, it is also desirable for a model to achieve accurate total reserves by correctly modelling the individual cells, $X_{i,j}$. 
 
\textbf{Results were analysed qualitatively and quantitatively}. Qualitative analysis allowed the MDN's strengths and weaknesses to be located graphically, while quantitative analysis provided a more objective measure of the model's accuracy. The qualitative analysis conducted included the following plots:
\begin{enumerate}
\item \textbf{Central Estimates:} Plots of the MDN and ccODP's central estimates $\hat{\mu}_{i,j}$ were compared to actual losses of the dataset $X_{i,j}$, as well as the empirical mean calculated from hundreds of simulations of the same dataset. 
\item \textbf{Risk Margins:} Plots of the MDN and ccODP's mean-centred risk margins (at the 25\%, 75\% and 95\% level) were compared to empirical risk margins. 
\item \textbf{Total reserves:} The distributions of total reserves estimated by the MDN and ccODP, $\hat{R}$, were plotted alongside the empirical distribution of total reserves. 
\end{enumerate}

When analysing individual cells, $X_{i,j}$, the RMSE, Log Score and Quantile Score statistics were calculated on each loss triangle involved in the modelling. For total reserves, $R$, the MDN and ccODP was fit on \textbf{50 triangles for each of the four simulated data environments, and 10 independent triangles randomly partitioned from the AUSI dataset}, to generate reserve estimates for each triangle, $\hat{R}_i$ for $i= 1,2,3..50$. Let $\textbf{X} = \{ X_{i,j}: i + j > 41 \} $,  $\textbf{R} = \{R_i, i = 1,2,3,..50\}$, $f_{\hat{\textbf{X}}} = \{f_{\hat{X}_{i,j}}: X_{i,j}\in \textbf{X}   \}$ and $X_q$ be the $q^{th}$ quantile estimate of the variable $X$. The quantitative metrics used are calculated as follows:

\begin{description}
\item[1. Distributional forecast accuracy, using the log score metric (\eqref{eq:LogScore})]: 
\begin{equation}
LogScore(\textbf{X},f_{\hat{\textbf{X}}}) = \frac{\sum_{(i,j): X_{i,j} \in \textbf{X}}ln(f_{\hat{X}_{i,j}}(X_{i,j}))}{|\textbf{X}|} 
\label{eq:LogScore}
\end{equation}
A higher log score is desirable as it indicates a more accurate distributional fit for the lower triangle. The log score wasn't calculated when analysing total reserves, as the fitted distributions usually fell completely outside the simulated empirical distribution, setting the likelihood to 0.

\item[2. Central estimate forecast accuracy, using the RMSE metric (\eqref{eq:RMSE1}, \eqref{eq:RMSE2})]:
\begin{equation}
 RMSE(\textbf{X},{\hat{\textbf{X}}}) = \sqrt{\frac{\sum_{(i,j): X_{i,j} \in \textbf{X}} (X_{i,j} - \hat{X}_{i,j})^2  }{|\textbf{X}|} }
 \label{eq:RMSE1}
\end{equation}
\begin{equation}
 RMSE(\textbf{R},{\hat{\textbf{R}}}) = \sqrt{\frac{\sum_{i = 1}^{D} (R_i - \hat{R_i})^2  }{D} } \label{eq:RMSE2}  
\end{equation}
A lower RMSE indicates more accurate central estimates for the lower triangle and total reserves.

\item[3. Quantile forecast accuracy (75\% and 95\%), using quantile scores (\eqref{eq:qs1}, \eqref{eq:qs2} )]:
\begin{equation}
QS(\boldsymbol{\hat{X}_{q}}, \textbf{X}) = \frac{\sum_{(i,j): X_{i,j} \in \textbf{X}} (\mathbf{1}(X_{i,j}  < \hat{X}_{i,j,q}) - q)( \hat{X}_{i,j,q} - X_{i,j})  }{|\textbf{X}|}
\label{eq:qs1}
\end{equation}
\begin{equation}
QS(\boldsymbol{\hat{R}_{q}}, \textbf{R}) = \frac{\sum_{i = 1}^{D} (\mathbf{1}(X_{i,j}  < \hat{X}_{i,j,q}) - q)(  \hat{X}_{i,j,q} - X_{i,j})  }{D}
\label{eq:qs2}
\end{equation}
A lower quantile score indicates more accurate quantile estimates, for individual cells and total reserves. 
\end{description}

\section{Results}
\label{sec:Results}
In this section, we analyse the results of the MDN, with the ResMDN separately analysed in Section \ref{sec:practical}.
\subsection{Stable forecasts - rolling origin model validation}

\begin{figure}[htb]
\centerline{\includegraphics[width = 10cm]{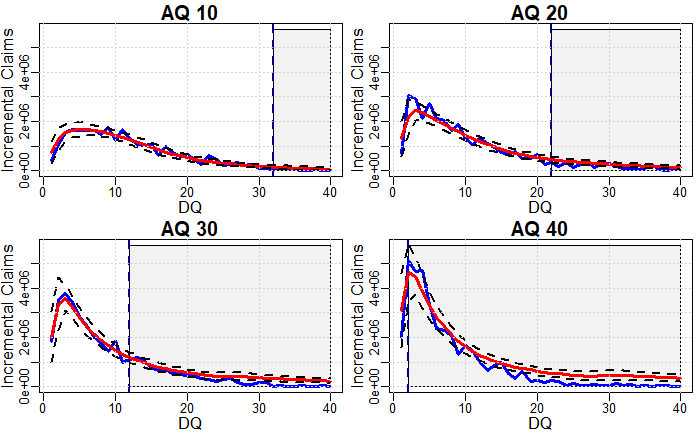}}
\caption{Environment 2 (speed up in claim processing): Plots of the mean (red) and standard deviation (black dotted) estimates of the MDN against actual losses (blue). The grey area represents the lower triangle, the forecasting region. These plots show the MDN producing reasonable and accurate forecasts, which were consistently observed.}
\label{fig:D2plotResults}
\end{figure}

Generally, the set of hyper-parameters selected by the rolling origin method produced very reasonable and accurate central and distributional forecasts. All models were successful in predicting a decrease in the mean and volatility of claims in the later development quarters (DQs), which is a significant achievement given the low quantity of data available to the MDN in those periods. Figure \ref{fig:D2plotResults} plots the MDN's mean and volatility estimates on environment 2, showing the accuracy of projections despite its systematic complexity.

\begin{figure}[htb]
\centerline{\includegraphics[width = 10cm]{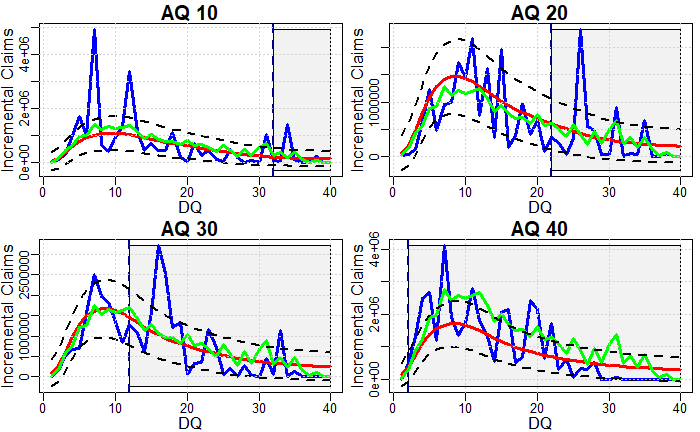}}
\caption{Environment 4: Plots of the MDN's central estimate (red) and standard deviation (dotted black) fits against the ccODP model (green) fit and actual losses (blue). The grey area represents the lower triangle, the forecasting region. These plots demonstrate the smooth and robust forecasts produced by the MDN, especially relative to the ccODP, despite the volatile data given.}
\label{fig:D4plotComparison}
\end{figure}

The MDN also produced smooth, robust predictions. This can especially be seen in Figure \ref{fig:D4plotComparison}, which plots the MDN's and ccODP's fits to the highly volatile environment 4. The ccODP's coefficients are derived from few data points, especially in the later AQs and DQs, leading to a more volatile fit throughout. Meanwhile, the MDN produced a more holistic fit, resulting in a smoother more robust forecast.

\begin{figure}[htb]
\centerline{\includegraphics[width = 10cm]{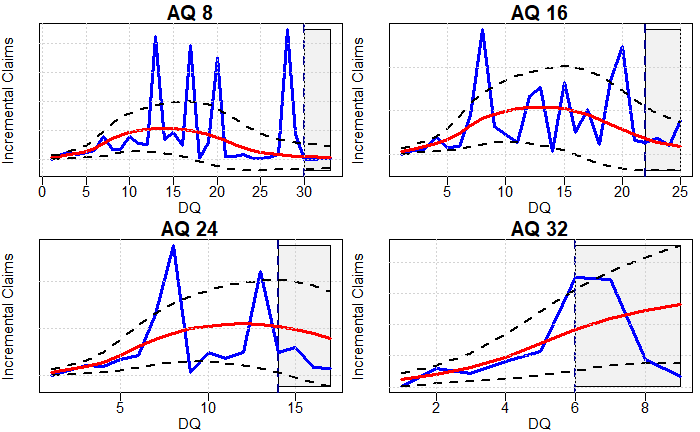}}
\caption{AUSI: Plotting the MDN's central estimate (red) and standard deviation (black dotted) fits against actual losses (blue). The grey area represents the lower triangle, the forecasting region. The MDN provides smooth and accurate forecasts using real data.}
\label{fig:AUSIplotResults}
\end{figure}

The rolling origin method, in the third partition, uses the latest calendar quarters for validation. A model that overfits the data will not project accurately, and hence the MDN is encouraged to produce a smooth fit. In addition to Figure \ref{fig:D4plotComparison}, the smoothness can be visualised in Figure \ref{fig:AUSIplotResults}, where the MDN produces a significantly smooth fit despite the huge volatility present in the dataset. 

The rolling origin model validation method proved successful at partitioning triangles of a size as small as 36$\times$36. The scarcity of data relative to the large datasets to which neural networks are usually applied  would normally discourage the use of this method. However, both the MDN and rolling origin partition performed well on less than 700 data points, showing their appropriateness in a practical loss triangle reserving setting.   
However, as suspected, the rolling origin method was visibly unable to capture the inflation shock accurately for environment 3, due to a lack of training data in the later calendar periods. This shortcoming was detected through analysis of the residuals in the upper triangle, which were consistently negative in the high inflation periods. Hence, an adjusted data partition methodology was implemented for environment 3, visualised in Section \ref{app:partition}, which allocated more training data to the later calendar periods, allowing the MDN more exposure to the inflation shock. This adjustment enabled the MDN to capture the later trends more effectively, and is recommended for data sets where a significant change in the claim pattern is observed in later calendar periods. 

\clearpage

\begin{figure}[htb]
\centerline{\includegraphics[width = 10cm]{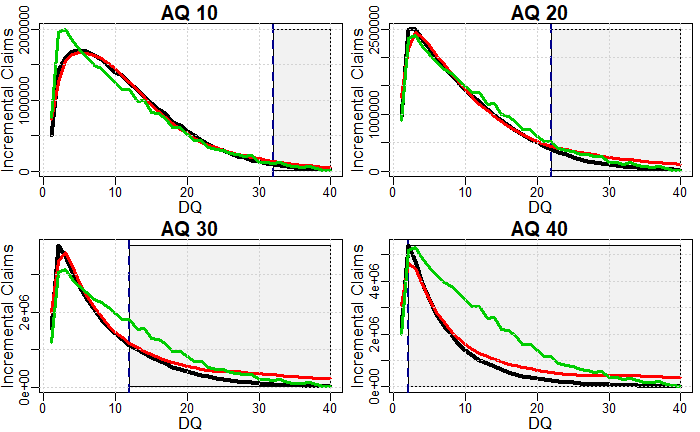}}
\caption{Environment 2: Plots comparing the mean estimates of the MDN (red) and ccODP (green) models to the empirical mean claims based on 250 simulations (black). The grey area represents the lower triangle, the forecasting region. The MDN captured the increase in claim settlement speed, while the ccODP did not.}
\label{fig:D2plotReal}
\end{figure}

\begin{figure}[htb]
\centerline{\includegraphics[width = 10cm]{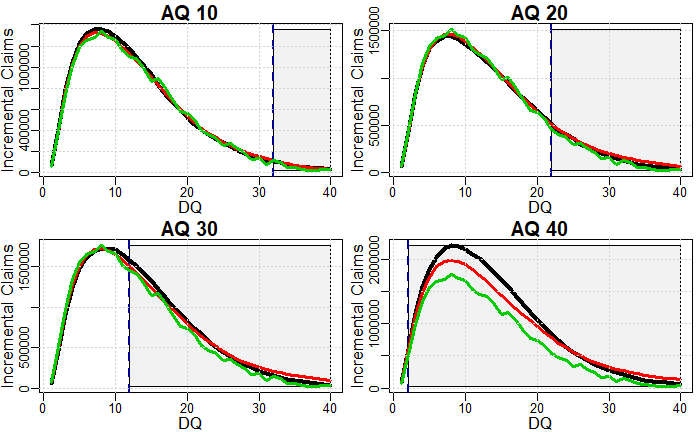}}
\caption{Environment 3: Plots comparing the mean estimates of the MDN (red) and ccODP (green) models to the empirical mean claims based on 500 simulations (black). The grey area represent the lower triangle, the forecasting region. The MDN captured the inflation shock more accurately than the ccODP.}
\label{fig:D3plotReal}
\end{figure}

\subsection{Distributional forecasting with the Mixture Density Network}

The Mixture Density Network (MDN), overall, outperformed the ccODP for all environments in all qualitative and quantitative metrics. When analysing both individual cells and total reserves, the MDN outperformed the ccODP, often in a decisive manner. 

Generally, the MDN's out-performance relative to the ccODP is due to its high flexibility, being able to fit a highly flexible function to the data and capture non-linearities. This flexibility can easily lead to over-fitting, but the rolling origin model validation method ensured that the MDN's flexibility was enough to capture the relevant trends in the data while minimising over-fitting. 

\subsubsection{Central estimate analysis}

\begin{figure}[htb]
\centerline{\includegraphics[width = 10cm]{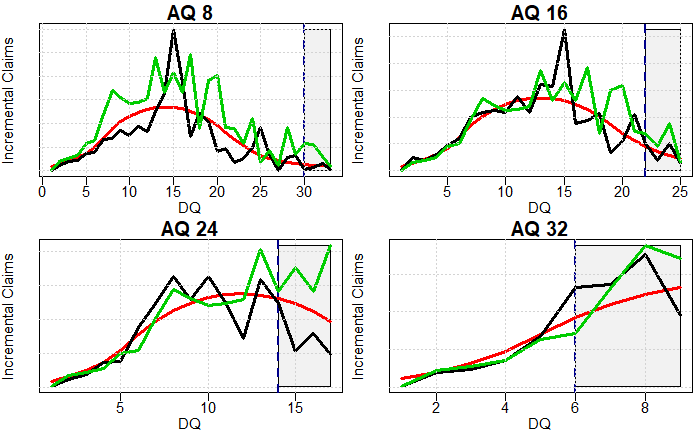}}
\caption{AUSI: Plots comparing the mean estimates of the MDN (red) and ccODP (green) models to the empirical mean claims based on 10 triangles (black). The grey area represent the lower triangle, the forecasting region. These plots show the MDN producing fairly accurate mean forecasts in a real environment, outperforming the ccODP.}
\label{fig:AUSIplotReal}
\end{figure}

The MDN produced excellent central estimate projections in all environments. Where the environment had more structural heterogeneity, i.e. where the ccODP assumptions are not satisfied, the MDN decisively outperformed the ccODP in all metrics. Several key observations can be made:
\begin{itemize}
\item In environment 1, the data (by design) satisfies the ccODP assumptions well, hence the ccODP was very competitive. Nevertheless, the MDN slightly outperformed in all quantitative metrics when measuring the accuracy of incremental claims, $X_{i,j}$. This can be attributed to the smooth function fit by the MDN. 
\item In environment 2, the MDN successfully learned that claims processing speed is increasing, predicting a sharper spike in claim payments in the later AQs. Figure \ref{fig:D2plotReal} plots the results for this environment. The ccODP, assuming homogeneity in claim development, approximated claims as medium-tailed, leading to a clear over-estimation of claims in later AQs.
\item For environment 3, the MDN accurately captured the inflation shock at calendar quarter 30 (CQ30) onwards. Figure \ref{fig:D3plotReal} plots the results. The ccODP did not keep pace with the increased inflation due to its limited ability to handle heterogeneity, leading to its under-estimation of claims from CQ30 onwards. 
\item In both environment 4 and the AUSI environment, the MDN handled volatile data well and provided accurate central estimates, outperforming the ccODP. This accuracy is shown in Figure \ref{fig:AUSIplotReal}, which plots the MDN's central estimates against the empirical mean based on 10 simulations. 
\end{itemize}

\begin{figure}[htb]
\centerline{\includegraphics[width = 9cm]{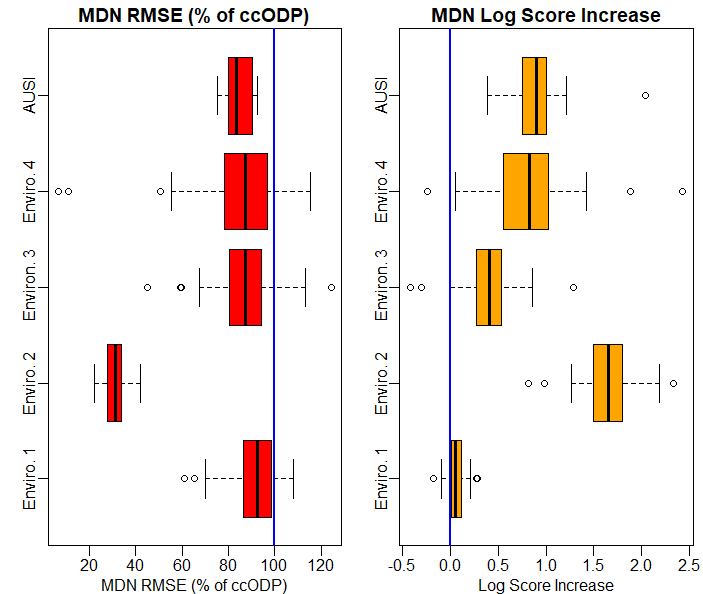}}
\caption{
\textbf{Left:} conventional boxplots displaying the MDN's RMSE as a percentage of ccODP's for each of the 50 triangles run for environments 1,2,3,4, and 10 triangles for AUSI.
\textbf{Right:} conventional boxplots displaying the MDN's increase in log score relative to the ccODP for each of the 50 triangles run for environments 1,2,3,4, and 10 triangles for AUSI. }
\label{fig:RMSE}
\end{figure}

Figure \ref{fig:RMSE} provides boxplots of the MDN's \% reduction of the RMSE relative to the ccODP for 50 triangles in each of the environments tested (10 for the AUSI environment). The boxplots show that the MDN achieved a positive reduction in the RMSE (lower RMSE) relative to the ccODP for the majority of triangles in each environment, further showing the MDN's higher forecasting accuracy. 

Despite the MDN's success, it showed weaknesses in several areas, some of which were mitigated:
\begin{itemize}
\item Using the NLL loss function alone can encourage the MDN to over-estimate the volatility when its central estimate is inaccurate. This was seen in environment 2, where a MDN with a NLL loss function under-estimates claims in the (40,2) cell, leading to an excessively high volatility estimate for that region. In addition to sigma activity regularisation (see Section \ref{sec:algorithm}, adding an MSE term to the loss function helped to resolve this issue, as it encouraged the MDN to achieve more accurate central estimates. Hence, an MSE term was added to the loss function for environments 1 and 2, and is recommended for loss triangles with sharp shifts in claims development.
\item Taking the log of aggregate claims linearised the data, which often led to faster and more accurate modelling. In this paper, environments 1 and 3 were fit with a mixture Log-Gaussian, as they showed significantly more accurate results, especially capturing the claims decay in later DQs. 

\end{itemize}

\begin{figure}[htb]
\centerline{\includegraphics[width = 10cm]{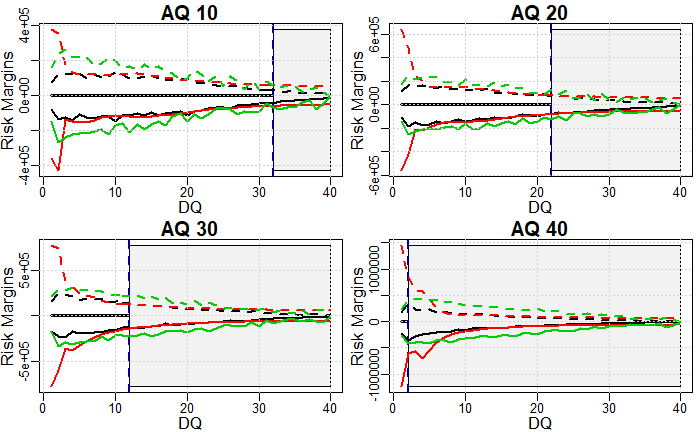}}
\caption{Environment 2: Plots comparing the 25\% (solid) and 75\% (dashed) risk margin estimates of the MDN (red) and ccODP (green) models to the empirical margins based on 250 simulations (black). The grey area represents the lower triangle, the forecasting region. These plots demonstrate the MDN providing more accurate volatility forecasts than the ccODP benchmark.}
\label{fig:D2plotShape}
\end{figure}
\begin{figure}[htb]
\centerline{\includegraphics[width = 10cm]{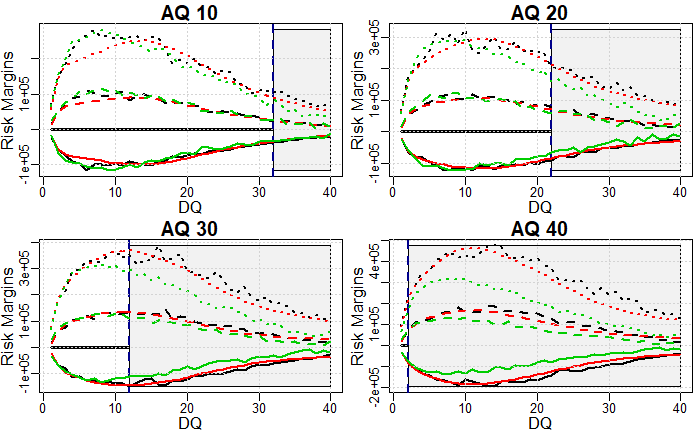}}
\caption{Environment 3 (Inflation Shock): Plots comparing the 25\% (solid), 75\% (dashed) and 95\% (dotted) risk margin estimates of the MDN (red) and ccODP (green) models to the empirical margins based on 500 simulations (black). The grey area represents the lower triangle, the forecasting region. Similarly to Figure \ref{fig:D2plotShape}, this figure demonstrates the MDN's ability to capture the increase in claim volatility due to the inflation shock, while the ccODP did not.}
\label{fig:D3plotShape}
\end{figure}

\clearpage
 
\subsubsection{Volatility estimate analysis}

The Mixture Density Network produced very smooth and accurate volatility estimates. Where noise in the data was low, the MDN projected low volatility, and vice versa. Overall, it outperformed the ccODP qualitatively and quantitatively in estimating the volatility of individual cells. 

\begin{figure}[htb]
\centerline{\includegraphics[width = 10cm]{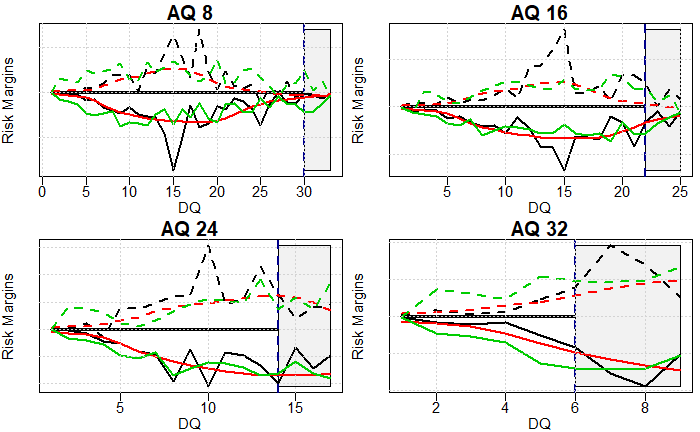}}
\caption{AUSI: Plots comparing the 25\% (solid) and 75\% (dashed) risk margin estimates of the MDN (red) and ccODP (green) models to the empirical margins based on 10 triangles (black). The grey area represents the lower triangle, the forecasting region. This figure demonstrates the MDN producing accurate and smooth volatility forecasts on real data.}
\label{fig:AUSIplotShape}
\end{figure}

In relation to individual cells, the MDN's risk margin estimates at the 25th and 75th percentile were more accurate overall than the ccODP's margins in almost all environments tested. The ccODP's variance is a function of its mean, and hence it failed where the central estimates failed. For example, in environment 2 (Figure \ref{fig:D2plotShape}), the ccODP over-estimates claims in later AQs, which led to it over-estimating margins in that same period. In environment 3, the ccODP under-estimated claims in later calendar Quarters (CQs), as it didn't effectively capture the inflation shock. This led to volatility estimates also being too low in those periods, as Figure \ref{fig:D3plotShape} illustrates. The MDN dealt with these structural issues more effectively, leading to more accurate dispersion estimates of claims. In the volatile AUSI environment, the MDN produced smooth and accurate margin estimates, as Figure \ref{fig:AUSIplotShape} illustrates. 

While some correlation was found in the results between central and volatility forecasts, the MDN allows a large degree of independence between the $\mu$ and $\sigma$ parameters estimated, allowing it to fit a much wider range of distributions than the ccODP. 

Figure \ref{fig:RMSE} provides boxplots of the MDN's increase in the log score relative to the ccODP for 50 triangles in each of environments 1,2,3 and 4. The AUSI boxplot was based on 10 triangles. The boxplots show that the MDN achieved a higher log score relative to the ccODP for the majority of triangles in each environment, indicating a more accurate probabilistic forecast. 

With the MDN's success, there are some weaknesses which must be addressed:
\begin{itemize}
\item The MDN still showed signs of attributing noise to systematic trends. In environment 2, even though the DQ2 spike was fixed, the volatility was still too high for AQ30 and AQ40.
\end{itemize}
\begin{itemize}
\item Similar to central estimates, the MDN often over-estimates volatility in later DQs, also due to a lack of data in that region. 
\end{itemize}

\clearpage

\subsubsection{Quantile estimate analysis}

The MDN provided more accurate 75\% and 95\% quantiles for all environments in the majority of triangles run for each environment. These results follow from the MDN's ability to provide more accurate central estimates and volatility estimates. The quantile analysis was mainly quantitative, using the quantile scores. Table \ref{table:mean} confirms that in all environments, the MDN reduces the 75\% and 95\% quantile scores for the majority of triangles, indicating more accurate quantile estimates at the 75\% and 95\% levels. 

The MDN and ccODP models were run on fifty triangles of each of environments 1,2,3,4, and the ten triangles partitioned from the AUSI data. The quantitative metrics are calculated for the 50 triangles and averaged, with results between the MDN and ccODP models compared in Table \ref{table:mean}. As the table shows, the MDN, on average, had a lower RMSE and Quantiles Scores and had a higher log score for each environment, which is a significant out-performance by the MDN. Table \ref{table:triangles} further reinforces these results by showing the percentage of triangles in which the MDN outperformed the ccODP for each quantitative metric. In each environment, the MDN outperforms the ccODP in each metric for the majority of triangles.

\begin{table}[h!]
\centering
\begin{tabular}{|c ||c|c|c| c| c| c |} 
 \hline
 Environment & Model & Mean RMSE & RMSE  & Mean LS & Mean QS   & Mean QS \\ 
 &&&(\% of ccODP)&&(75\%)&(95\%) \\ [0.5ex] 
 \hline\hline
  1 & ccODP& 1,656,921.0&100 & -14.99 & 380,968.5 & 144,423.7  \\
 \hline
 \textbf{1} & \textbf{MDN} &\textbf{ 1,527,799} &\textbf{92.2}& \textbf{-14.93} & \textbf{375,413.6} & \textbf{140,754.3} \\ 
 
 \hline\hline
 2 & ccODP& 591,505.4 &100& -14.97 & 111,733.3  & 31,213.2 \\
 \hline
\textbf{2} & \textbf{MDN}& \textbf{182,041.1} &\textbf{30.8}& \textbf{-13.31} & \textbf{50,628.9} &\textbf{ 16,326.1}  \\ 
%\hline
%2 & ResMDN & 378,898.3 &64.1& -15.23 & 81,641.2 & 30,300.2 \\
 \hline\hline
 3& ccODP& 190,482.4 &100& -13.46 & 57,562.2 & 32,164.1  \\
 \hline
  \textbf{3} & \textbf{MDN} & \textbf{162,621.8} &\textbf{85.4}& \textbf{-13.05} & \textbf{47,837.0} & \textbf{18,817.4} \\ 

 \hline\hline
 4 & ccODP& 1,053,008.0 &100& -14.72 & 232,011.3  & 96,506.1 \\
 \hline
 \textbf{4} & \textbf{MDN}&\textbf{ 652,230.7} &\textbf{61.9}& \textbf{-13.88} & \textbf{210,694.1} & \textbf{95,235.0}  \\

 \hline\hline
AUSI & ccODP& - &100& -14.04 & 124,976.3  & 58,760.6 \\
 \hline
  \textbf{AUSI} & \textbf{MDN}&\textbf{ -} &\textbf{84.2}& \textbf{-13.07} & \textbf{105,559.9} & \textbf{53,773.9}  \\
 \hline
\end{tabular}
\caption{The average score, over 50 triangles, of each quantitative metric; the RMSE, log score (LS) and quantile scores (QS) for the 75\% and 95\% levels. The MDN outperformed the ccODP in all environments and metrics when the average is taken. 
}
\label{table:mean}
\end{table}
\begin{table}[h!]
\centering
\begin{tabular}{ |c||c|c|c|c|c|  }
 %\hline
 %\multicolumn{6}{|c|}{Triangles where ccODP is Out-Performed (\%)} \\
 \hline
 Environment& Model & RMSE &Log Score&Quantile Score (75\%)& Quantile Score (95\%)\\
 \hline\hline
 1   & MDN & 76    &80&   60 & 58\\
 \hline\hline
 2& MDN&   100  & 100   &100& 100\\
%\hline
% 2   & ResMDN & 96    &58&   90 & 72\\
 \hline\hline
3 & MDN&88 & 94&  90& 100\\
 
 \hline\hline
 4    & MDN&84 & 98&  66& 50\\

 \hline\hline
 AUSI &MDN   &100 & 100&  100& 90\\
  \hline
\end{tabular}
\caption{The percentage of triangles in which the MDN outperformed the ccODP for each environment and metric.
}
\label{table:triangles}
\end{table}

\subsubsection{Total reserves}

\begin{figure}[htb]
\centerline{\includegraphics[width = 10cm]{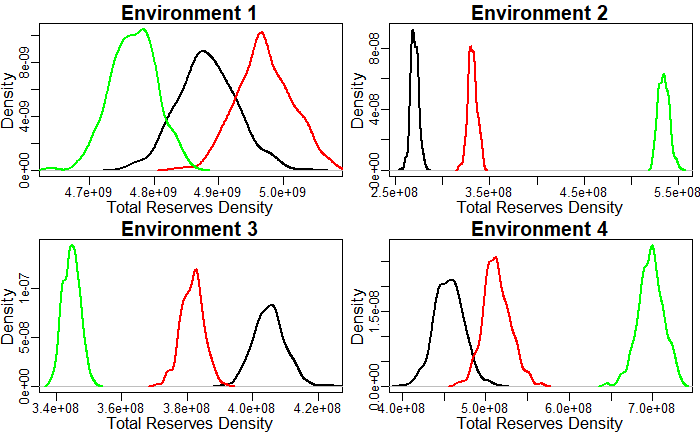}}
\caption{A plot of the total reserve density estimates for all environments, $\hat{R}$, showing the MDN (red) and ccODP's (green) estimated densities against the empirical density (black) based on hundreds of simulations. For each environment, only one triangle is analysed for each plot. The MDN consistently provides more accurate results, except for environment 1  }
\label{fig:TotalReserves}
\end{figure}

The MDN, in all environments except environment 1, showed more accurate central and dispersion estimates of total reserves compared to the ccODP estimate (the dispersion accuracy is qualitatively measured through visualising reserve density plots, Figure \ref{fig:TotalReserves} plots these results). An empirical distribution of total reserves, based on many simulations, is used as an estimate of the actual distribution of $R$. This out-performance follows from the MDN modelling the mean and volatility of individual cells more accurately than the ccODP. Because the claims in environment 1 are homogeneous in development, the ccODP provides highly competitive results and the MDN didn't outperform. However, for the more complicated environments 2,3 and 4, the MDN had more accurate 75\% and 95\% quantiles of total reserves compared to the ccODP. Table \ref{table:totalreserves} calculated the quantitative metrics for both models for total reserve estimates, $\hat{R}$. The qualitative analysis (Figure \ref{fig:TotalReserves}) also supports these results. 

\begin{table}[h!]
\centering
\begin{tabular}{|c ||c| c| c| c |} 
 \hline
 Environment & Model & RMSE ($\times 10^6$)&   QS(75\%) ($\times 10^6$)  & 
  QS(95\%) ($\times 10^6$)\\ [0.5ex] 
 \hline\hline
 1 & MDN & 111.7 & 37.3 & 15.4 \\ 
 \hline
 \textbf{1} & \textbf{ccODP}&  \textbf{92.9}  &\textbf{28.9} & \textbf{13.2}  \\
 \hline\hline
 \textbf{2} & \textbf{MDN}& \textbf{80.4}  & \textbf{20.1} &\textbf{4.25}  \\
 %\hline
 %2 & ResMDN&  101.1  &27.0 & 13.69  \\
 \hline
 2 & ccODP& 260.2  & 65.9  & 13.45 \\
 \hline\hline
  \textbf{3} & \textbf{MDN} & \textbf{39.5}  & \textbf{19.7} & \textbf{20.00} \\ 
 \hline
 3& ccODP& 53.0  & 37.0 & 44.37  \\
 \hline\hline
 \textbf{4} & \textbf{MDN}&\textbf{99.4}  & \textbf{24.1} & \textbf{9.00}  \\
 \hline
 4 & ccODP& 322.8  & 56.5  & 11.98 \\
 \hline
\end{tabular}
\caption{The RMSE and quantile scores (QS) at the 75\% and 95\% levels, calculated for total reserve estimates, $\hat{R}$. The ccODP outperforms for Environment 1, but the MDN outperforms otherwise.}% The ResMDN outperforms the ccODP for Environment 2 in terms of the RMSE and 75th quantile.}
\label{table:totalreserves}
\end{table}

%\clearpage

\section{Practical Considerations }
\label{sec:practical}

\subsection{Projection constraints}
\label{sec:Projection}

\begin{figure}[htb]
\centerline{\includegraphics[width = 10cm]{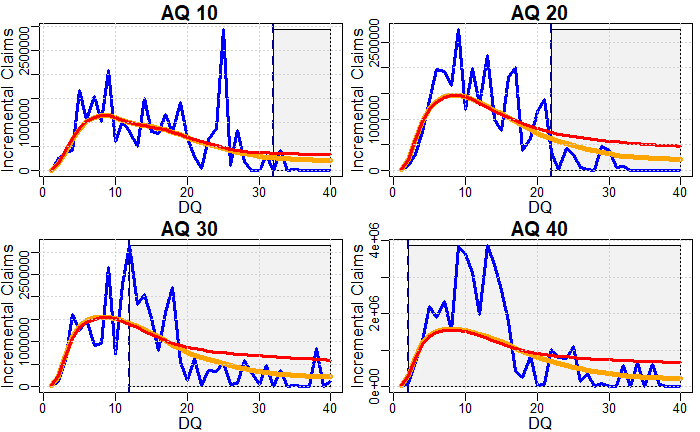}}
\caption{Environment 4: A plot of the central estimates forecasted by the MDN (red) and MDN with projection constraints (orange), against the actual losses (blue). The grey area represents the lower triangle, the forecasting region. This figure demonstrates the MDN producing more reasonable results when projections are constrained. }
\label{fig:D4ECN}
\end{figure}

For both the MDN and ResMDN, central estimate projections in the lower triangles can be explicitly constrained for any desired set of cells, $X_{i,j}$. Without constraints these projections can, and sometimes do, produce negative results in individual cells, and even for total reserves for some accident years (as was seen for the ResMDN in Section \ref{sec:ResMDN}). This is unrealistic in some circumstances, and is best prevented by constraining projections to non-negativity. Similarly, it may be desirable to force projected payments to converge toward zero with increasing DQ. There might be other “reasonableness constraints” that the actuary wishes to apply. In this paper, projection constraints were applied successfully to the ResMDN for environments 2 and 3. Mean forecasts were constrained in the later DQs to be non-negative. There are some points to note:
\begin{itemize}
\item The MDN's forecast was virtually unchanged in the upper triangle, meaning the constraints set did not distract the model from fitting the in-sample data accurately. 
\item The MDN produced a natural, smooth curve while still meeting the constraints. There was no evidence of a sudden jolt in the fitted function, nor did the function simply rest on the closest constraint boundary. Consequently, only a small proportion of cells need to be constrained to achieve reasonable results. For example, the non-negativity constraint described above was only applied to approximately 10\% of cells in the lower triangle.
\end{itemize}
 To visually demonstrate the effect of constraining projections, we apply this methodology in environment 4, shown in Figure \ref{fig:D4ECN}. The volatile data in this environment caused the MDN to occasionally over-estimate losses in the later DQs. Hence, the mean was constrained in that period to be approximately 0, which the model followed once set. In general, the actuary can set any desirable boundary to any cell in the lower triangle, to ensure the MDN strongly leans towards fitting functions with sensible projections. %A potentially useful application would be to take a GLM benchmark and constrain the network's mean estimates to remain within a certain percentage of the GLM. This would allow the user more control of the output by restricting deviation from a known model while still utilising the network's modelling power (to an extent).

\subsection{Interpretability: ResMDN} \label{sec:ResMDN}

The ResMDN shows plenty of potential in boosting the residuals of its GLM backbone while providing more interpretable results than the MDN. In this paper, we analyse the ResMDN on environments 2 and 3, as in both environments, the ccODP provided a smooth backbone with clear and detectable flaws which could be analysed. In both environments, the ResMDN successfully detected the ccODP's shortcomings and corrected them, to an extent. Both mean and volatility estimates were corrected; the ResMDN didn't just boost the mean, but also the distribution. A visible shortcoming of the ResMDN is that it can produce unreasonable forecasts; for environment 2 the model occasionally under-estimates claims in later DQs, while for environment 3 there were some instances where it over-estimated the inflation shock. Hence we apply projection constraints in the later DQs to mitigate these deviations.

\begin{figure}
\centering
\begin{subfigure}{.5\textwidth}
  \centering
  \includegraphics[width=0.95\linewidth]{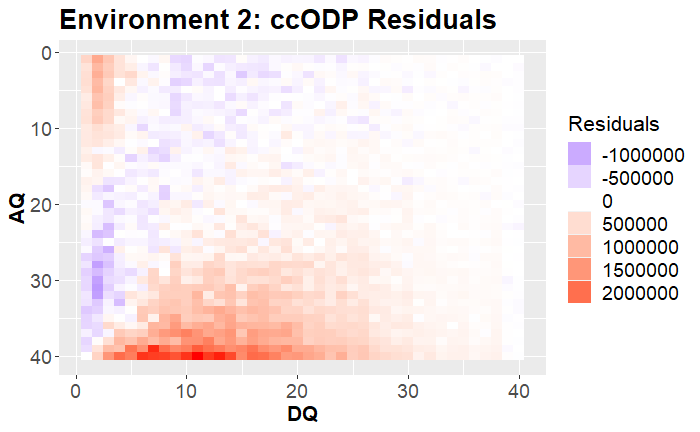}
  \caption{A heatmap of the ccODP's residuals}
  \label{fig:map1}
\end{subfigure}%
\begin{subfigure}{.5\textwidth}
  \centering
  \includegraphics[width=0.95\linewidth]{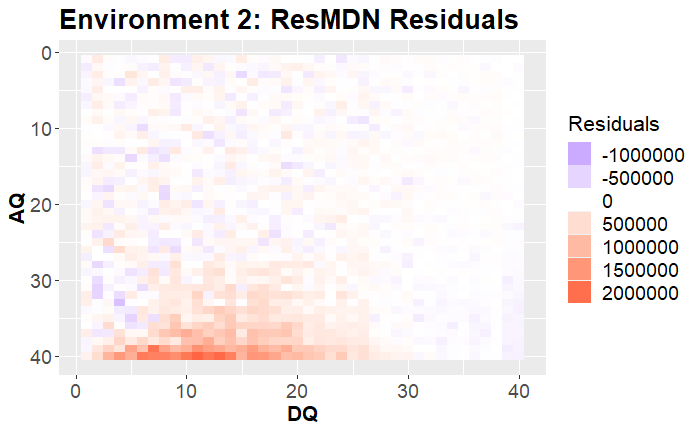}
  \caption{A heatmap of the ResMDN's residuals}
  \label{fig:map2}
\end{subfigure}
\caption{Environment 2: Heatmaps showing the ccODP's initial residuals in (a), calculated as $\mu_{i,j}^{ccODP} - X_{i,j}$. The ResMDN's residuals, calculated as $\mu_{i,j}^{ResMDN} - X_{i,j}$, are shown in (b). The lighter colours in (b) show that the ResMDN partially corrected the ccODP's residuals, producing more accurate forecasts.  }
\label{fig:ResMDNheatmap}
\end{figure}

\begin{figure}[htb]
\centerline{\includegraphics[width = 10cm]{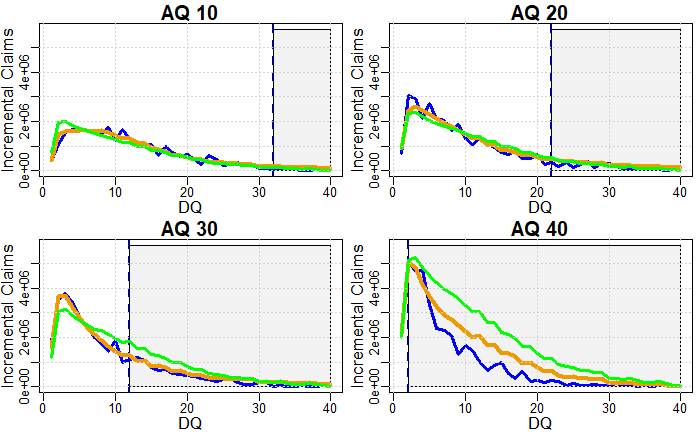}}
\caption{Environment 2: Plots comparing the mean estimates of the ResMDN (orange) and ccODP (green) models to actual losses (blue). The grey area represents the lower triangle, the forecasting region. These plots demonstrate the ResMDN partially correcting the ccODP's residuals, leading to more accurate forecasts.}
\label{fig:ResMDNplotReal}
\end{figure}

\begin{figure}[htb]
\centerline{\includegraphics[width = 10cm]{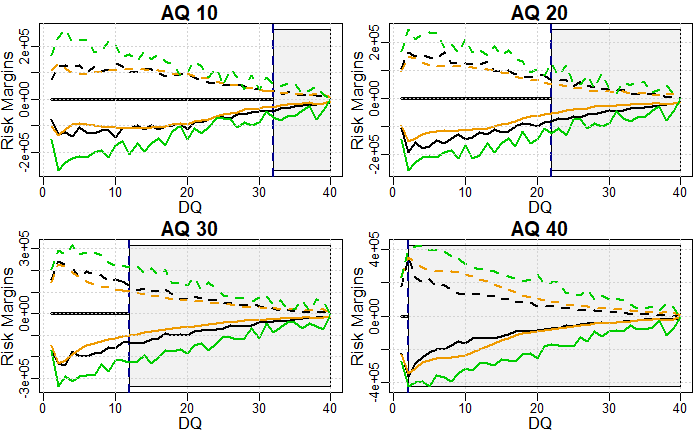}}
\caption{Environment 2: Plots comparing the 25\% (solid) and 75\% (dashed) risk margin estimates of the ResMDN (orange) and ccODP (green) models to the empirical margins based on 250 simulations (black). The grey area represents the lower triangle, the forecasting region. While Figure \ref{fig:ResMDNplotReal} demonstrates the ResMDN's ability to correct the ccODP's mean estimates, this figure shows the ResMDN correcting the ccODP's volatility estimates, allowing distributional forecasts to be more in line with empirical data.}
\label{fig:ResMDNplotShape}
\end{figure}

\begin{figure}[htb]
\centerline{\includegraphics[width = 10cm]{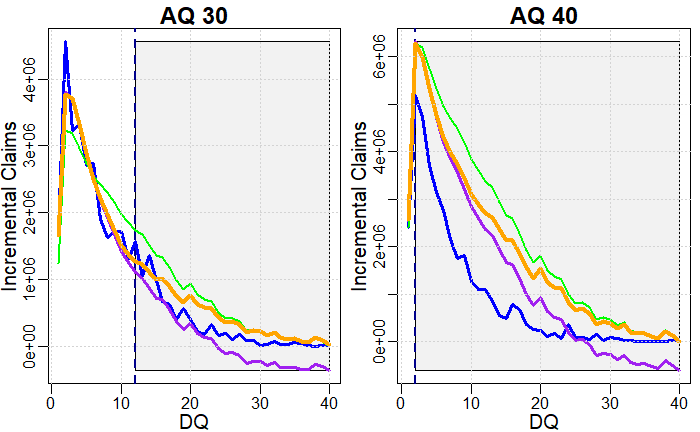}}
\caption{Environment 2: Plots comparing the mean estimates of the constrained (orange) and unconstrained (purple) ResMDN models against the ccODP backbone (green) and actual losses (blue). The grey area represents the lower triangle, the forecasting region. These plots show that the unconstrained ResMDN can significantly under-estimate claims, hence justifying the use of constraints to stabilise forecasts.}
\label{fig:ResMDNconst}
\end{figure}

\begin{itemize}
\item The ResMDN demonstrated the ability to recognise errors in the ccODP's central estimates and correct them. In environment 2, where the claim processing speed gradually increases, the ccODP models the claims as being middle-tailed, as it assumes homogeneity in claim development. Hence, the ccODP under-estimated claims in early AQs and over-estimated claims later on. Figures \ref{fig:map1} and \ref{fig:map2} show heatmaps of the ccODP and ResMDN's residuals for environment 2, respectively. As can be seen, the ResMDN successfully understood the ccODP's shortcomings just mentioned, and reduced the residuals. Figure \ref{fig:ResMDNplotReal} plots the central estimates of the ResMDN and ccODP models, also showing the ResMDN's corrections producing a more accurate forecast. 
\item The ResMDN also demonstrated the ability to recognise errors in the ccODP's volatility estimates and correct them. In environment 
2, the ccODP models the speed up in claims processing with a higher dispersion parameter, leading to consistently over-estimating volatility. The ResMDN  successfully learns this shortcoming and reduces volatility estimates accordingly. Figure \ref{fig:ResMDNplotShape} visualises the ResMDN's boosting for environment 2; the ResMDN has corrected volatility to almost match the empirical trend, indicating its higher distributional forecast accuracy. 

\end{itemize}
The ResMDN showed promise in detecting and partially correcting the embedded GLM's mean and volatility estimates. Given these results, several considerations arose that are worth mentioning:
\begin{itemize}
\item The ResMDN projected the ccODP's residuals unreasonably in some instances. For example, in environment 2, it learnt that the ccODP over-estimates claims in later AQs, so it adjusted this error by reducing the loss estimates for those periods, and continued that correction excessively in the later DQs. This issue was fixed by constraining projections for DQ 38-40. Central estimate bounds for these periods were set between 0-500,000 (for environment 2) and 0-200,000 (for environment 3). These bounds are judgemental, but are reasonably wide to accommodate the uncertainty in forecasting. Figure \ref{fig:ResMDNconst} plots the difference in forecasts between the constrained and unconstrained ResMDN. While the plotted triangle is an extreme scenario (out of the 50 triangle sample), it justifies using the constrained ResMDN to ensure incremental claims tend to 0.
\item Despite the ResMDN correcting residuals, its log score is disadvantaged compared to the ccODP. This is mainly driven by the high coefficient of variation of incremental claims observed in the later DQs (of which only a small amount are in-sample), which naturally favoured the ODP distribution over the mixture Gaussian.

\end{itemize}

Tables \ref{table:meanresmdn}, \ref{table:trianglesresmdn} and \ref{table:totalreservesresmdn} display the quantitative results of the unconstrained ResMDN (ResMDN) and constrained ResMDN (ResMDN-PC), in a similar fashion to Tables \ref{table:mean}, \ref{table:triangles} and \ref{table:totalreserves}. In accordance with conclusions from visual analysis in Figures \ref{fig:ResMDNplotReal} and \ref{fig:ResMDNplotShape}, the ResMDN produced more accurate central estimates, 75th and 95th quantiles when looking at individual cells, demonstrating that it has improved the mean and distribution over its ccODP backbone. In the majority of triangles, the ResMDN achieved a more accurate RMSE and Quantile Score. Boosting residuals translated to more accurate total reserve estimates, for the mean, 75th and 95th quantiles. We note that the unconstrained ResMDN yielded more accurate central estimates for environment 2 than the constrained model. This is due to the constraints slightly discouraging boosting to avoid negative forecasts. Moreover, the apparent improvement in performance embodied in the reduced RMSE can come at the cost of negative projections.

\begin{table}[h!]
\centering
\begin{tabular}{|c ||c|c|c| c| c| c |} 
 \hline
 Environment & Model & Mean RMSE & RMSE  & Mean LS & Mean QS   & Mean QS \\ 
 &&&(\% of ccODP)&&(75\%)&(95\%) \\ [0.5ex] 
 \hline\hline
 
2 & ResMDN & 378,898.3 &64.1& -15.23 & 81,641.2 & 30,300.2 \\
\hline
2 & ResMDN-PC & 436,414.9 &73.8& -14.95 & 78,584.0 & 21,644.8 \\
\hline
3 & ResMDN & 190,471.4 &100.0& -14.01 & 55,497.3 & 25,348.4 \\
\hline
3 & ResMDN-PC & 182,944.4 &96.0& -13.65 & 53,003.2 & 26,971.7 \\
 \hline\hline
 
\end{tabular}
\caption{The average score, over 50 triangles, of each quantitative metric; the RMSE, log score (LS) and quantile scores (QS) for the 75\% and 95\% levels for the ResMDN. The ResMDN-PC model features constrained projections
}
\label{table:meanresmdn}
\end{table}

\begin{table}[h!]
\centering
\begin{tabular}{ |c||c|c|c|c|c|  }
  \hline
 Environment& Model & RMSE &Log Score&Quantile Score (75\%)& Quantile Score (95\%)\\
 \hline\hline

 2   & ResMDN & 96    &58&   90 & 72\\

  \hline
  2   & ResMDN-PC & 96    &60&   96 & 100\\

  \hline
   3   & ResMDN & 78    &38&   80 & 76 \\ %72\\
   \hline
   3   & ResMDN-PC & 90    &46&   92 & 76 \\
 \hline\hline
\end{tabular}
\caption{The percentage of triangles in which the ResMDN outperformed the ccODP in each metric. The ResMDN-PC model features constrained projections
}
\label{table:trianglesresmdn}
\end{table}

\begin{table}[h!]
\centering
\begin{tabular}{|c ||c| c| c| c |} 
 \hline
 Environment & Model & RMSE ($\times 10^6$)&   QS(75\%) ($\times 10^6$)  & 
  QS(95\%) ($\times 10^6$)\\ [0.5ex] 
 \hline\hline
 
 2 & ResMDN&  101.1 (39\%)  &27.0 (41\%) & 13.69 (102\%)  \\
 \hline
2 & ResMDN-PC&  180.2 (69\%)  &44.85 (68\%) & 9.13 (68\%)  \\
 \hline
 3 & ResMDN&  55.9 (105\%)  &19.9 (54\%) & 18.3 (41\%) \\%16.9  \\
\hline
 3 & ResMDN-PC&  36.7 (69\%) &22.3 (60\%)& 25.8 (58\%) \\%16.9  \\
 \hline\hline
\end{tabular}
\caption{The RMSE and quantile scores (QS) at the 75\% and 95\% levels, calculated for total reserve estimates, $\hat{R}$. The number in brackets represents the corresponding metric as a $\%$ of the corresponding ccODP results. The ResMDN-PC model features constrained projections}
\label{table:totalreservesresmdn}
\end{table}

\subsubsection{Comparison to the MDN}

While the ResMDN was able to detect the ccODP's systematic errors, it generally failed to outperform the non-embedded MDN in the above examples as it only partially corrected the errors (see Figure \ref{fig:ResMDNheatmap} for an illustration). Nevertheless, the ResMDN showed great function and potential in correcting an embedded GLM's mean and volatility estimates, while maintaining a fair level of interpretability. While the current implementation is not as accurate as the MDN, if this is criteria is essential then one possibility is to consider a more sophisticated GLM backbone. 

Finally, it should also be noted that an additional benefit of the ResMDN is that the ResMDN's training time was noticeably faster than the MDN. For environment 2, when a few triangles were analysed, the ResMDN finished training in 2218 epochs on average, compared to 3868 for the MDN. This 43\% reduction in training time is due to the ResMDN's GLM initialisation being more accurate than the MDN's random starting fit, meaning that less parameter adjustment is required. This observation is in parallel to the findings of \citet*{GaRiWu2020}, who also noted faster training times under the CANN structured model.

\section{Conclusion} \label{sec:conclusion}
In this paper, we identified, addressed, and mitigated a number of obstacles, which so far hindered the popularisation of neural networks in loss triangle reserving. A neural network design which specialises in distributional forecasting, the MDN, was applied successfully to a variety of environments. The MDN produced more accurate central and volatility estimates, for both the individual cell and total reserve measures. The rolling origin model validation method provided a framework for model testing and selection, suited to sequential data. This sequential data partition gave preference to smooth, robust models, while also producing accurate forecasts. 

We considered additional extensions involving projection controls and hybrid GLM-MDN approaches. We demonstrated that the MDN was able to significantly outperform the ccODP model in a variety of environments and metrics. While the ccODP model is not representative of the full potential of GLMs in loss reserving, it is a gold standard, and as such the results are compelling.

\section*{Acknowledgements}

Earlier versions of this paper were presented at the Actuaries Institute 2021 Virtual Summit, and at the ASTIN Online Colloquium. The authors are grateful for constructive comments received from colleagues who attended those events. 

This research was supported under Australian Research Council's Linkage (LP130100723, with funding partners Allianz Australia Insurance Ltd, Insurance Australia Group Ltd, and Suncorp Metway Ltd) and Discovery (DP200101859) Projects funding schemes. The views expressed herein are those of the authors and are not necessarily those of the supporting organisations. 

\section*{Data and Code}
We are unable to provide the dataset that was used in the empirical case study due to confidentiality. However, simulated data sets with similar features, as well as all relevant codes, can be found at \url{https://github.com/agi-lab/reserving-MDN-ResMDN}.

\section*{References}

\bibliographystyle{elsarticle-harv}
\bibliography{2021refs}

\appendix

\section{Fitting a mixture Log-Gaussian} \label{app:log}

\begin{lemma}
Let $Y$ be a strictly positive random variable. Suppose that $X=ln(Y)$ follows a mixture Gaussian distribution with parameters $(\boldsymbol{\alpha}, \boldsymbol{\mu}, \boldsymbol{\sigma})$, such that:
$$  f_{X}(x) = \sum_{k = 1}^{K}\alpha_k \phi( {x | \mu_k, \sigma_k}  ).   $$
Then $Y$ follows a mixture Log-Gaussian distribution with parameters $(\boldsymbol{\alpha}, \boldsymbol{\mu}, \boldsymbol{\sigma})$, such that:
$$  f_{Y}(y) = \frac{1}{y}\sum_{k = 1}^{K}\alpha_k   \phi(ln(y) | \mu_k, \sigma_k). $$
\end{lemma}
\noindent\textbf{Proof:} 
By definition $X$ is defined by \[X=\{X_i \textrm{, w.p } \alpha_i, i=1,..,K\},\] where $0<\alpha_i<1$ and $\sum_{i=1} ^K \alpha_i=1$. Then for some function $g(\cdot)$, if we consider $Y=g(X)$, we then have by definition \[Y=\{g(X_i) \textrm{, w.p } \alpha_i, i=1,..,K\}.\] i.e. in the log-gaussian case (where we have $g(\cdot)$ being the exponential function) we have $Y$ defined as a mixture of $g(X_i)$, which are individually log-gaussian.

\section{Training Data in the Later Calendar Periods} \label{app:partition}
\begin{figure}
\centering
\subfloat[Partition 1: Testing models]{\includegraphics[width = 0.5\linewidth]{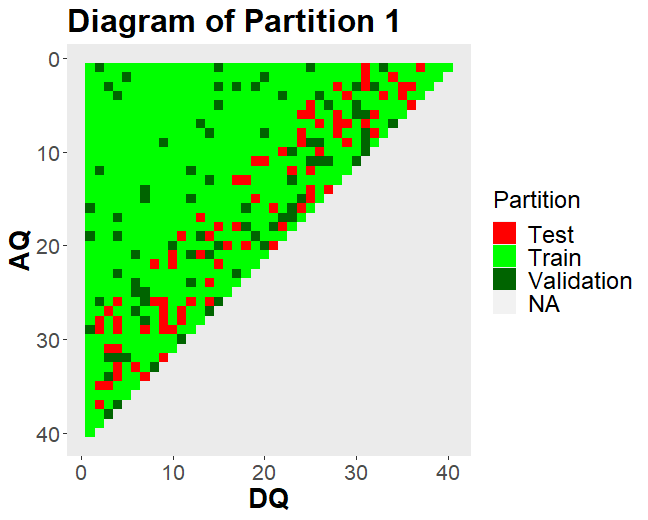}} 
\subfloat[Partition 2: Testing models with alternate testing set]{\includegraphics[width = 0.5\linewidth]{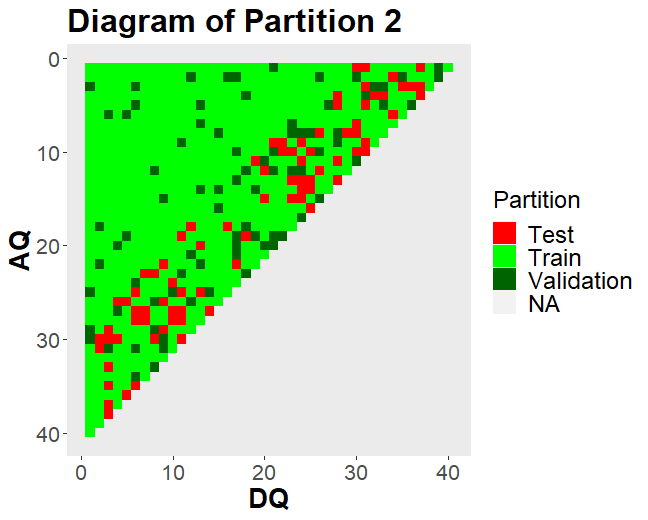}}\\
\subfloat[Partition 3: Fitting the final model]{\includegraphics[width = 0.5\linewidth]{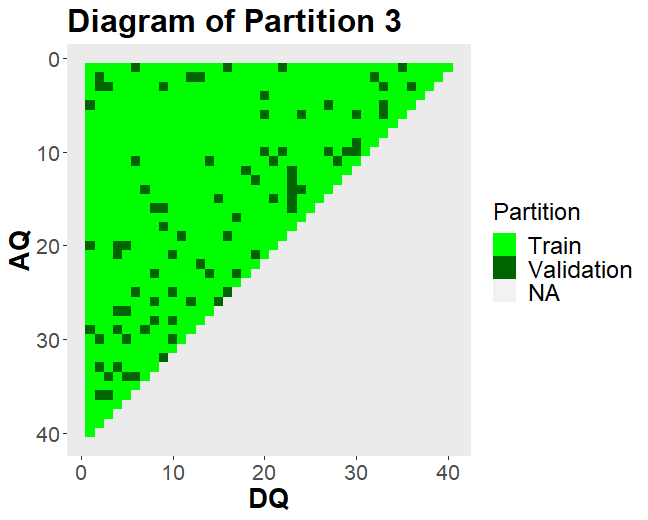}}
\subfloat[Partition 4: Fitting the final model with alternate validation set]{\includegraphics[width = 0.5\linewidth]{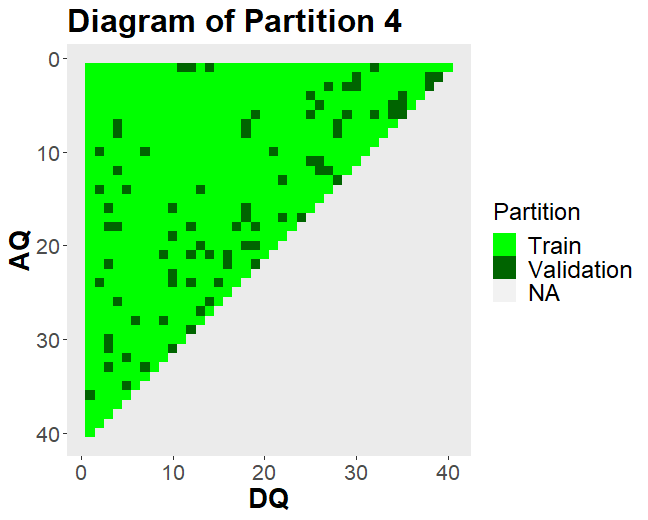}} 
\caption{Diagrams of the four data partitions used when the latest calendar quarters were included in the training set}
\label{fig:4parts}
\end{figure}

For environments where a significant shift in claim development pattern is observed in the later calendar periods (e.g. in environment 3, and observable in the generated data), the rolling origin method is less likely to capture that shift. Hence, an adjusted data partition is implemented in such situations, as Figure \ref{fig:4parts} illustrates. The focus remains on ensuring the validation and testing sets are in the later calendar periods, but also includes some training data there to help the MDN model the shift. The partition is made as such:
\begin{itemize}
\item All partitions used the entire upper triangle
\item 10\% of the data was assigned to each of the validation and testing sets
\item For Partition 1, the testing set was randomly assigned in the latest 11 calendar quarters. Then, half the validation set was assigned to non-testing points in the latest 11 calendar quarters, while the rest of the validation set was randomly assigned to the whole triangle. 
\item For Partition 2, the process for Partition 1 was repeated, except the testing set excluded any data points that were in the test set for Partition 1. This allowed both testing sets to cover a wider variety of data points.
\item For Partitions 3 and 4, no testing set was assigned as the final model was trained on them. The validation set was assigned similarly to Partitions 1 and 2, in that half the set was chosen randomly from the latest 11 calendar quarters, while the other half was chosen randomly from the entire triangle.
\item The final model was trained five times; three times under Partition 3 and twice using Partition 4
\end{itemize}

\section{Final Model Designs}

Table \ref{table:models} lists the most accurate model design found by the algorithm in Section \ref{sec:algorithm}.
\begin{table}[h!]
\centering
\begin{tabular}{ |c|c|c|c||c|c|c|c|c|c|  }
 \hline

 \hline
 Environment& Model &Component Distribution& MSE weight& $\lambda_w$ &$\lambda_\sigma$&$p$ & $n$ &$h$ & $K$\\
 \hline \hline
 1   & MDN &Log-Gaussian&3&0&0.0001&0&60&4&2\\
 \hline \hline
 2& MDN&Gaussian&3&0.001&0.1&0.1&100&2&4\\
 \hline
 2   & ResMDN &Gaussian&4&0&0&0&20&2&4\\
 \hline \hline
3 & MDN&Log-Gaussian&0&0&0.0001&0&80&3&1\\
 \hline
 3   & ResMDN &Gaussian&4&0*&0&0.1&60&2&2\\
 \hline \hline
 4    & MDN&Gaussian&0&0&0&0&40&4&3\\
 \hline  \hline
 AUSI &MDN&Gaussian&0   &0&0&0.1&20&3&3\\
  \hline
\end{tabular}
\caption{The most accurate model designs determined by the algorithm listed in Section \ref{sec:algorithm}. $\lambda_w$ represents the L2 weight regularisation coefficient, $\lambda_\sigma$ represents the L2 activity regularisation coefficient on the $\sigma$ output, $p$ is the dropout rate, $n$ is the number of neurons in each hidden layer, $h$ is the number of hidden layers and $K$ is the number of components in the mixture density. 
*The weight regularisation coefficient was manually adjusted to 0 in order to encourage boosting. 
}
\label{table:models}
\end{table}

\section{Additional Notes on Modelling }
\label{App:additional}
While effort was taken to keep the modelling as methodical and simple as possible, there were some additional details worth mentioning regarding the methodology in this paper:
\begin{itemize}
\item The ResMDN was trained with an MSE term (with a weight of 4) to encourage the model to boost mean estimates. 
\item The number of components for the ResMDN was pre-determined, set as the number of components deemed best for the \textbf{mixture Gaussian} MDN on the same environment.
\item For environments 1 and 3, fitting mixture Log-Gaussians gave visibly more accurate results than fitting a mixture Gaussian, hence that distribution was chosen for these environments.
\item An MSE term (with a weight of 3) was added to environments 1 and 2, to assist in capturing the peak in claims in the later development quarters. 
\item The LogScore for each individual cell was restricted to a minimum of $-50$. The statistic was often heavily influenced by low-volume data points in the later DQs, which tended to score lower due to their high coefficient of variation. As the log score in the lower triangle is usually between -8 and -20, this restriction predominantly applies to these low volume data points and reduces their influence on the score. 
\item When processing the fit of a mixture Log-Gaussian distribution, the $\sigma_k$ for $k = 1,2,..K$ was set to a maximum of $2\sqrt{Var(\textbf{X})}$, where $\textbf{X} = \{X_{i,j}, i + j \leq 41\}$. This was done to avoid high $\sigma$ values which, when transformed from mixture Gaussian to mixture Log-Gaussian, would lead to unreasonably high volatility estimates.
\item The hyper-parameter selection algorithm was only run on a single triangle from each environment. The chosen model for that triangle was used to fit all 50 triangles of that environment. This was done to increase the efficiency of modelling. 
\item When running the ResMDN on environment 3, a positive L2 penalty on the weights led to almost no boosting. Hence, the penalty was manually adjusted to 0 to encourage network activity. 
\item Out of the 210 triangles the MDN and ResMDN were fit on, the results displayed in Section \ref{sec:Results} generally aim to illustrate the average results obtained. Of course, some triangles yielded better results, while others yielded worse results. 

\end{itemize}

\end{document}